\documentclass[aps,prb,reprint]{revtex4-1}
\usepackage{amsmath,amssymb} % mathematics
\usepackage{amsmath} % mathematics
\usepackage{amsfonts} % mathematics
\usepackage{bm} % bold in math-mode
\usepackage{here}
\usepackage{url}
\usepackage[pdftex]{graphicx} % figure
\usepackage{braket} % braket
\usepackage{physics}
\begin{document}
%%%%%%%%%%%%%%%%%%%%%%%%%%%%%%%%%%%%%%%%%%%%%%%%%%%%%%%%%%%%
%%%%%%%%%%%%%%% title and author information %%%%%%%%%%%%%%%
%%%%%%%%%%%%%%%%%%%%%%%%%%%%%%%%%%%%%%%%%%%%%%%%%%%%%%%%%%%%
% Title
\title{Emergent SU(3) magnons and thermal Hall effect in the antiferromagnetic skyrmion lattice}
% Author information
\author{Hikaru Takeda$^{1\star}$}
\email{takeda.hikaru@issp.u-tokyo.ac.jp}
\author{Masataka Kawano$^{2\star}$}
\email{masataka.kawano@tum.de}
\author{Kyo Tamura$^1$, Masatoshi Akazawa$^1$, Jian Yan$^1$}
\author{Takeshi Waki$^3$, Hiroyuki Nakamura$^3$}
\author{Kazuki Sato$^4$, Yasuo Narumi$^4$, Masayuki Hagiwara$^4$} 
\author{Minoru Yamashita$^1$}
\author{Chisa Hotta$^5$}
\affiliation{\vspace{4mm}\\$\:^1$Institute for Solid State Physics, University of Tokyo, Kashiwa, 277-8581, Japan}
\affiliation{$\:^2$ Department of Physics, Technical University of Munich, 85748 Garching, Germany}
\affiliation{$\:^3$Department of Materials Science and Engineering, Kyoto University, Kyoto 606-8501, Japan}
\affiliation{$\:^4$Center for Advanced High Magnetic Field Science (AHMF), Graduate School of Science, Osaka University, Toyonaka, Osaka 560-0043, Japan}
\affiliation{$\:^5$ Department of Basic Science, University of Tokyo, Meguro-ku, Tokyo 153-8902, Japan}

%%%%%%%%%%%%%%%%%%%%%%%%%%%%%%%%%%%%%%%%
%%%%%%%%%%%%%%% abstract %%%%%%%%%%%%%%%
%%%%%%%%%%%%%%%%%%%%%%%%%%%%%%%%%%%%%%%%
\begin{abstract} 
Complexity of quantum phases of matter is often understood by the underlying gauge structures, 
as was recognized by the $\mathbb{Z}_2$ and U(1) gauge theory description of spin liquid in frustrated magnets. 
Anomalous Hall effect of conducting electrons can intrisically arise from U(1) gauges expressing 
the spatial modulation of ferromagnetic moments or from SU(2) gauges representing the spin-orbit coupling effect. 
Similarly, in insulating ferro and antiferromagnets, 
the magnon excitations can contribute to anomalous transports 
by feeling the U(1) and SU(2) gauges arising from the features of ordered moments or interactions. 
In this work, we report the emergent higher rank SU(3) gauge structure in the magnon transport 
based on the thermal conductivity measurements of MnSc$_2$S$_4$ in an applied field up to 14\,T. 
The thermal Hall coefficient takes a substantial value when the material enters 
a three-sublattice antiferromagnetic skyrmion phase, 
which is confirmed by the large-scale spin wave theory. 
The excited magnons are dressed with SU(3) gauge field, which is a mixture of three species of U(1) gauge fields 
originating from the slowly varying magnetic moments on these sublattices. 
\end{abstract}
\maketitle
%%%%%%%%%%%%%%%%%%%%%%%%%%%%%%%%%%%%%%%%%%%%
%%%%%%%%%%%%%%% Introduction %%%%%%%%%%%%%%% 
%%%%%%%%%%%%%%%%%%%%%%%%%%%%%%%%%%%%%%%%%%%%
\par
%%%%%%%%%%%%%%%%%%%%%%%%%%%%%%%%%%%%%%%%%%%%%%%%%%%%%%%%%%%%%%%%%%%%%%%%%%%%%%%%%%%%%%%%%%%%%%%%%%%%%%%%%%%%%%%%%%%%%%%%
%%%%%%%%%%%%%%%%%%%%%%%%%%%%%%%%%%%%%%%%%%%%%%%%%%%%%%%%%%%%%%%%%%%%%%%%%%%%%%%%%%%%%%%%%%%%%%%%%%%%%%%%%%%%%%%%%%%%%%%%
\noindent
\section{Introduction}
Quantum phases of matter are very often complex and require mathematical ingenuity to clarify the nature of emergent phenomena. 
One famous example is the bound states of kinks that appear as gapped low energy excitations of 
the seemingly simple Ising spin system in CoNb$_2$O$_6$, 
which turned out to follow an emergent E$_8$ exceptional Lie algebra symmetry\cite{Coldea2010} 
based on the integrable field theory\cite{Zamolodchikov1989}. 
Indeed, there are several other cases that effective theories explaining the low-energy excitations 
of interesting quantum phases are not the simple bosonic or quasi-particle ones  
but are those subject to gauge fields. 
In the Kitaev model, the spin-1/2 degrees of freedom separate into Majorana fermions and fluxes of 
an emergent ${\mathbb Z}_2$ gauge field\cite{Kitaev2006}. 
The half-quantized thermal Hall conductivity reported in $\alpha$-RuCl$_3$ is 
argued as being carried by these fractionalized Majorana fermions\cite{Kasahara2018,MYamashita2020,Yokoi2021,Bruin2022}. 
In the pyrochlore systems, quantum fluctuations transform a spin ice state to a U(1) spin liquid phase 
characterized by the emergent lattice electrodynamics with U(1) global gauge symmetry. 
The masked pinch point singularities of inelastic scattering signature in 
Pr$_2$Zr$_2$O$_7$ is considered relevant to this state hosting a monopole excitation\cite{Kimura2013}. 
\par
When one refers to the gauges in material solids, they are quantum mechanical 
as there is always redundancy in the description of phases of wave functions of the related particles.  
While these gauges never break their symmetries in the way that the lattice symmetries mark the phase transitions, 
the related gauge invariants can play another important role. 
For example, the spin-orbit coupling of electrons or the non-coplanar structured localized magnetic moments 
can bear an emergent U(1) gauge field. 
This gauge field represents the fictitious magnetic field that bends the motion of charges  
and yields an anomalous Hall effect\cite{neubauer2009,lee2009,kanazawa2011,li2013}. 
There, the quantized Hall resistance is explained by the gauge invariant referred to as a Chern number. 
In chiral magnets such as MnSi\cite{Muhlbauer2009,Yu2010} that are also good metals, 
the conduction electrons feel an emergent gauge field as they travel through 
the spatially varying spin-textures and contribute to the topological Hall effect\cite{Tokura2007,Nagaosa2013}. 
\par
Magnons in insulating magnets are simple bosonic excitations but can also carry a U(1) gauge; 
the thermal Hall effect in ferromagnetic insulators\cite{onose2010} 
was the first to report the topological transport of magnons due to U(1) gauge field. 
Theories showed that the antisymmetric Dzyaloshinskii-Moriya(DM) spin exchange interactions or the non-coplanar 
structures of ordered moments can be represented by the gauge that bears the Berry curvature in the energy bands
\cite{fujimoto2009,katsura2010,matsumoto2011prb,matsumoto2011prl}. 
Furthermore, in the insulating versions of skyrmions, GaV$_4$Se$_8$ \cite{Akazawa2022}, 
the magnon thermal Hall effect is explained by the U(1) gauge fields\cite{Hoogdalem2013} similarly to the cases of metallic skyrmions.  
%*%*%*%*%*%*%*%*%*%*%*%*%*%*%*%*%*%*%*%*%*%*%*%*%*%*%*%*%*%
%*%*%*%*%*%*%*%*%*%*%*%*%*%*%*%*%*%*%*%*%*%*%*%*%*%*%*%*%*%
\begin{figure*}[tbp]
   \centering
   \includegraphics[width=17cm]{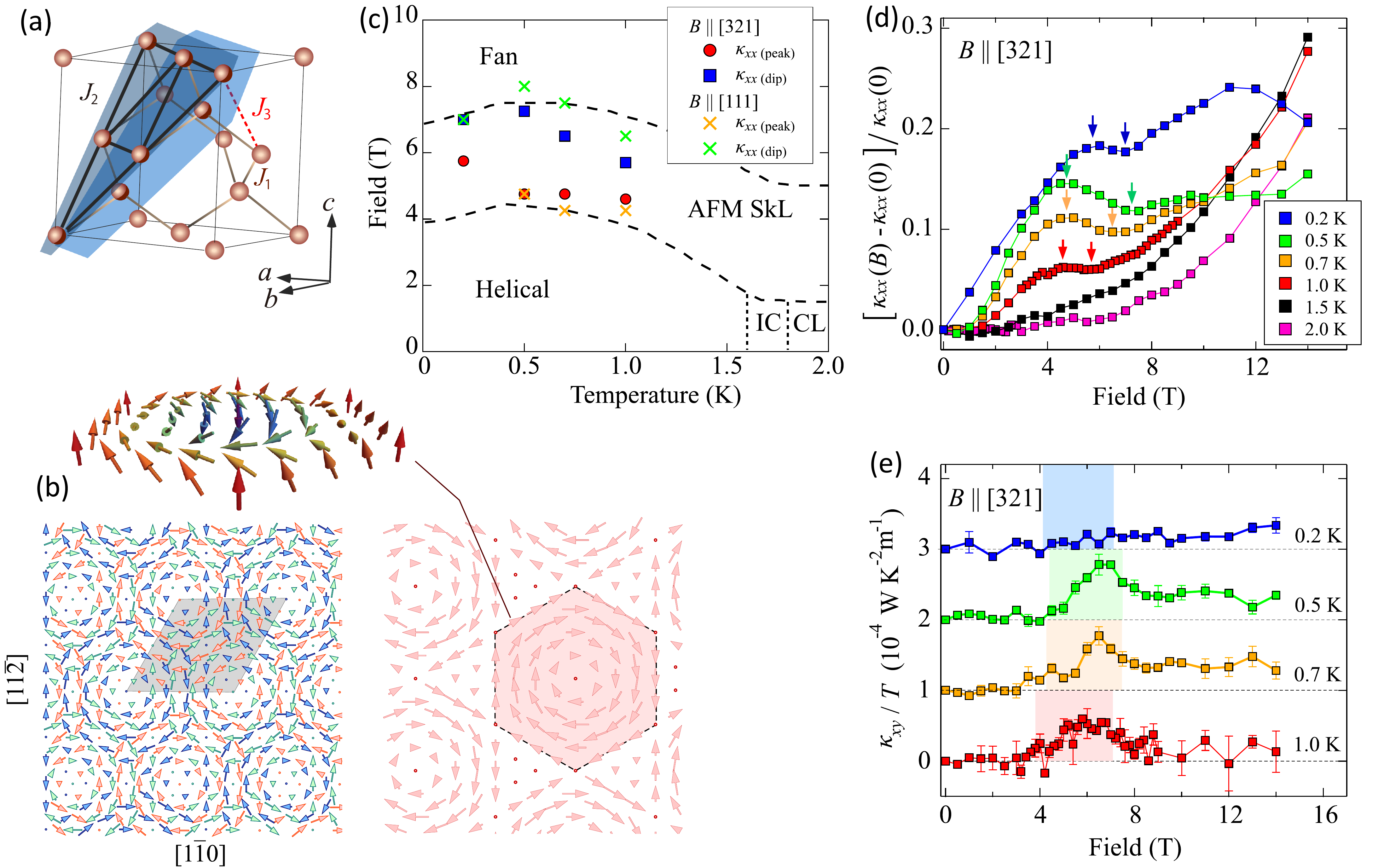}
   \caption{
(a) Diamond lattice formed by Mn$^{2+}$ ions (red circles) in MnSc$_2$S$_4$. 
The blue planes represent [111] planes where the Mn$^{2+}$ ions form a triangular lattice. 
The magnetic unit cell of AFM-SkL consists of six triangular layers. 
(b) Schematic figure of the AFM-SkL state in MnSc$_2$S$_4$ 
viewed along the [111] direction (left-bottom panel), 
where the spins on three different sublattices are shown in different colors, 
and the unit cell on that layer including 64 sites are shown. 
Spins on one of the sublattices forming a ferromagnetic skyrmion lattice 
are extracted on the right panel, and its hexagonal unit 
is shown in more detail on the top panel. 
(c) Magnetic field $B$ versus temperature $T$ phase diagram of MnSc$_2$S$_4$. 
Broken lines are phase boundaries determined by the neutron diffraction measurements\cite{Gao2020}. 
Filled circles and squares are the peaks and dips of $\kappa_{xx}$ for field direction 
$\bm B\parallel$[321], while crosses are those for $\bm B\parallel$[111]. 
(d) Field dependence of thermal conductivity $\kappa_{xx}$ normalized 
by its value at $B=0$\,T at $T\le 2$\,K. 
The arrows indicate the positions of peaks and dips, plotted in panel (c). 
(e) Field dependence of $\kappa_{xy}/T$ in the magnetically ordered phases. 
The shaded area represents the AFM-SkL phase \cite{Gao2017,Gao2020}. 
All the data points in $\kappa_{xy}/T$ are the ones averaged 
over the field-up and field-down measurements, 
and the error bars are the maximum deviations of the data from the averages. 
   }
\label{fig1}
\end{figure*}
%*%*%*%*%*%*%*%*%*%*%*%*%*%*%*%*%*%*%*%*%*%*%*%*%*%*%*%*%*%
%*%*%*%*%*%*%*%*%*%*%*%*%*%*%*%*%*%*%*%*%*%*%*%*%*%*%*%*%*%
\par
One may thus expect that a more abundant gauge structure may appear successively in the transport phenomena. 
However, even for simple two-sublattice insulating antiferromagnets in noncentrosymmetric crystals 
where the U(1) gauge picture is not applicable, 
it was only recently recognized that 
there can be another route using the SU(2) gauge field 
to have the anomalous thermal Hall effect\cite{kawano2019-1,kawano2019-2,kawano2019-3}. 
\par
Here, we report the experimental observation of the thermal Hall effect 
in the three-sublattice antiferromagnetic skyrmion lattice (AFM-SkL) realized in MnSc$_2$S$_4$. 
By performing a large-scale period spin-wave theory, 
we show that the carriers are the magnons in a complex SU(3) gauge field 
originating from the significant three-sublattice structure of skyrmions. 
\par
The AFM-SkL is a new class of skyrmion, recently discovered in a spinel compound MnSc$_2$S$_4$. 
As shown in Fig.~\ref{fig1}(a), the Mn$^{2+}$ ($S=5/2$) ions form a diamond lattice \cite{Fritsch2004,Reil2002} 
and undergo three successive magnetic transitions 
in a zero field, starting at $T\le T_N=2.3$\,K from a modulated collinear phase to an incommensurate phase 
and finally showing a helical magnetic long-range order below 1.6\,K \cite{Gao2017, Gao2020}. 
Interestingly, the phases above $T_N$ may not be a simple paramagnet but a classical spiral-spin liquid phase 
characterized by a highly degenerate manifold of states with a series of continuous wave numbers 
forming surface in the reciprocal space\cite{Bergman2007,Gao2017}. 
When a magnetic field $\bm B$ is applied along the [100], [110], and [111] directions, 
the helical phase transforms to the triple-$Q$ phase at around $B=4$--6\,T. 
Combining the neutron scattering experiment and Monte Carlo simulation, 
the triple-$Q$ state in $\bm B\parallel$[111] is identified as an AFM-SkL phase, 
which consists of three sublattices approximately forming a 120$^\circ$-types of antiferromagnetic order. 
The structure shown in Fig.~\ref{fig1}(b) is a cross-section of the diamond lattice 
forming a triangular lattice and stacks in the [111]-direction with 
six-fold periodicity (64 sites per triangular lattice layer, $N_c=384$ sites per unit cell). 
Each sublattice forms a triangular SkL, which can be well-explained theoretically 
within the classical framework\cite{Rosales2015, Rosales2022,Gobel2017}. 
%
%*%*%*%*%*%*%*%*%*%*%*%*%*%*%*%*%*%*%*%*%*%*%*%*%*%*%*%*%*%*%*%*%*%*%*%*%*%*%*%*%*%*%*%*%*
%*%*%*%*%*%*%*%*%*%*%*%*%*%*%*%*%*%*%*%*%*%*%*%*%*%*%*%*%*%*%*%*%*%*%*%*%*%*%*%*%*%*%*%*%*
\section{experimental}
\subsection{Synthesis and measurements} 
Single crystals of MnSc$_2$S$_4$ were synthesized by the chemical vapor transport method. 
We call two crystals with the shape of a thin plate with [321] plane as sample $\#$1 and [111] plane as sample $\#$2. 
The thermal-transport measurements were performed by the steady method by using a variable temperature insert (VTI) for 2--60 K and a dilution refrigerator (DR) for 0.1--3 K. 
The heat current $J_Q$ was applied along $[\bar{1}11]$ ($[2\bar{1}\bar{1}]$) and the magnetic field $\bm B$ was 
applied along [321] ([111]) for the sample $\#$1 ($\#$2). 
The detailed setup of the thermal transport measurements is shown in Appendix \S.\ref{ap:exp1}. 
In the main text, we show the experimental data for the sample $\#$1. 
We find that both samples show essentially the same field dependence (see section V in SM), 
indicating high reproducibility of our results, 
and also confirming that the two field directions yield the same magnetic phases. 
%%%%%%%%
%%%%%%%%
\subsection{Phase diagram}
\noindent
We performed thermal transport measurements on the single crystals of MnSc$_2$S$_4$ 
using the setup shown in Appendix \S.\ref{ap:exp1}. 
The detailed analysis of the field, temperature, and sample dependences are presented 
in Appendix \S.\ref{ap:exp-mfp}-\S.\ref{ap:exp-samp}. 
%%,ap:exp-kxx,ap:exp-kxy,ap:exp-samp}. 
Figure~\ref{fig1}(c) shows the $B$-$T$ phase diagram of MnSc$_2$S$_4$ at low temperature. 
The data points indicate the location of peaks and dips 
in the field dependence of the thermal conductivity ($\kappa_{xx}$) 
for two different field directions, $B \parallel [321]$ and $B \parallel [111]$, shown in Fig.~\ref{fig1}(d).
They are not much sensitive to the field direction, 
and show good agreement with the phase boundaries of the AFM-SkL phase 
obtained previously by the neutron diffraction experiment for $\bm B\parallel$ [111]\cite{Gao2020}. 
Although the temperature dependence of $\kappa_{xx}$ does not show a clear anomaly at $T_{N}$ 
(see Appendix Fig.~\ref{fSkxx_t}), 
the field dependence has distinct upturn and downturn, which are more visible for lower temperatures 
and dissapear at $T\gtrsim T_N$. 
The good correspondence of these features with the phase boundaries 
indicate that $\kappa_{xx}$ is strongly influenced by the magnetic ordering. 
%%%%%%%%
%%%%%%%%
\subsection{Thermal Hall measurement}
Figure \ref{fig1}(e) shows the field-dependence of 
thermal Hall conductivity $\kappa_{xy}/T$ at $T\le$ 1\,K for $\bm B\parallel$ [321] of MnSc$_2$S$_4$. 
In the helical phase at $B\lesssim$ 4\,T, $\kappa_{xy}/T$ is suppressed to nearly zero or slightly negative values, 
which, however, shows an abrupt and substantial increase in entering the AFM-SkL phase at around 4\,T. 
Its amplitude is overall suppressed at the lowest temperature 0.2\,K, 
which is usual for the thermal Hall conductivity that relies on thermally driven bosonic excitations. 
At above $\sim 8$\,T, where the previous theory predicts a fan phase, 
$\kappa_{xy}/T$ becomes suppressed but remain positive and finite. 

\subsection{Examination of the thermal conductivity}
\begin{figure}[tbp]
   \centering
   \includegraphics[width=7.5cm]{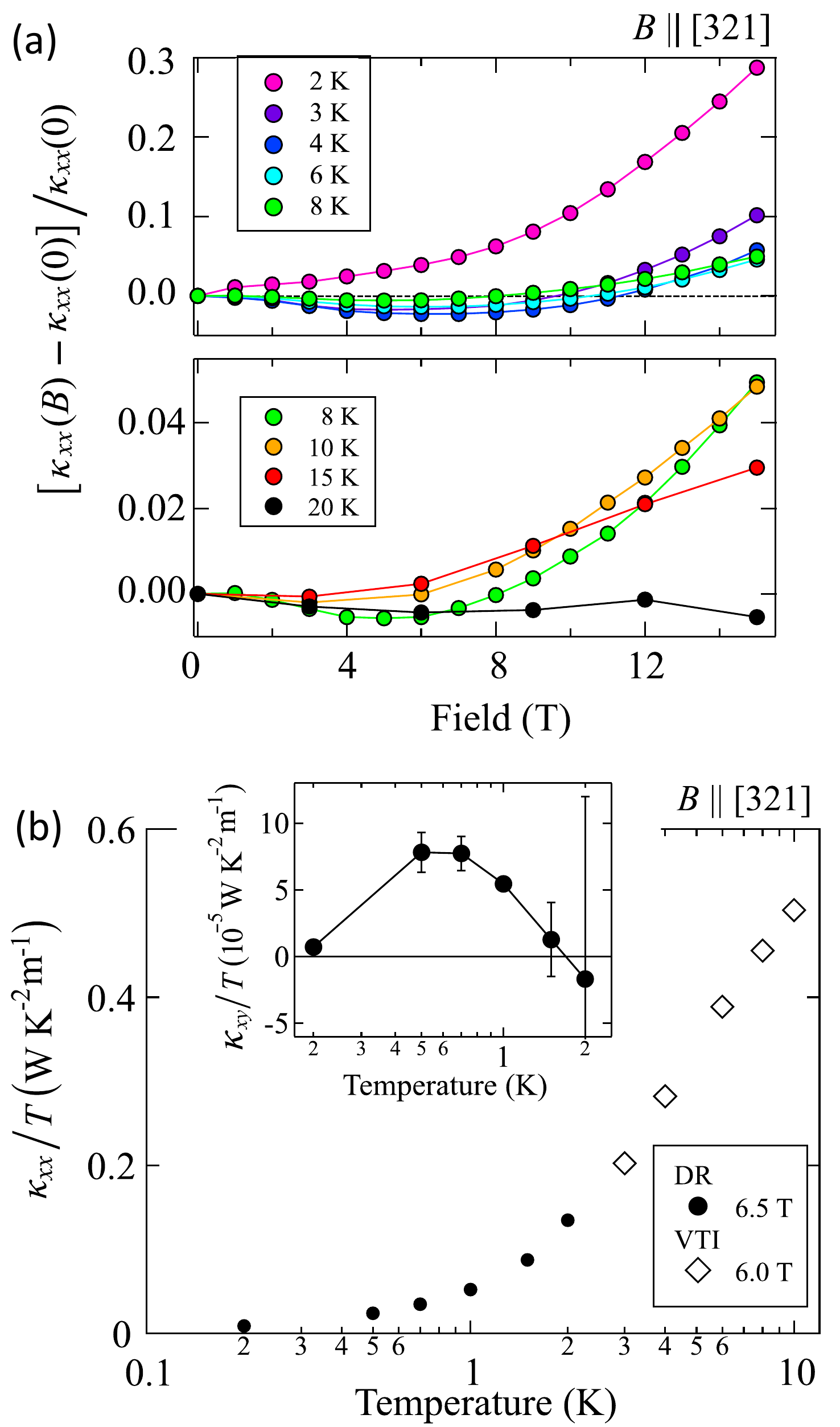}
   \caption{
(a) Field dependence of $\kappa_{xx}(B)$ normalized by the zero-field value above $T_N$.
(b) Temperature dependence of $\kappa_{xx}/T$ and $\kappa_{xy}/T$(inset) 
in the AFM-SkL phase (6.0 and 6.5\,T) 
obtained by the DR ($<$ 3\,K) and the VTI ($>$ 3\,K) measurements. 
   }
\label{fig3}
\end{figure}

In magnetic insulators, the carriers contributing to thermal transport can be phonons and magnons. 
Here, we clarify experimentally that $\kappa_{xx}$ has indeed a substantial contribution from magnons 
($\kappa_{xx}^{\rm mag}$)
on top of phonons ($\kappa_{xx}^{\rm ph}$). 
Figure~\ref{fig3}(a) shows the field dependence of $\kappa_{xx}$ at $T>T_N$ 
to be compared with Fig.~\ref{fig1}(d) at $T\le T_N$. 
The positive magnetothermal conductivity observed at lower fields for $T<T_N$ becomes negative above $T_N$. 
\par
The major effect of the magnetic field on the present system is to vary the magnon gap. 
In our theoretical calculation, when the system remains within the same phase 
the overall shape of the magnon bands does not change much while 
both the bandwidth and the gap vary (see Appendix Fig.~\ref{fStheory}). 
As shown in Fig.~\ref{fig2}(e), the magnon gap first decreases toward zero in approaching the 
helical-to-AFM-SKL phase transition point slightly below 4\,T, then increases again on entering the AFM-SkL phase, 
and closes at another transition point near 7\,T. 
In general, an decrease of magnon gap increases $\kappa_{xx}^{\rm mag}$ 
since the number of excited magnons that depends on the Bose distribution function increases. 
This can in turn suppress $\kappa_{xx}^{\rm ph}$, 
supposing that there is a sufficiently large amount of scattering of phonons by the magnons. 
\par
For this reason, the positive magnetothermal conductivity observed at $T<T_N$ in decreasing the gap 
inside the helical phase is the feature attributed to an increase of $\kappa_{xx}^{\rm mag}$. 
In the AFM-SkL phase, $\kappa_{xx}$ shows a slight decrease, 
which should be because of the re-opening of the magnon gap 
in addition to a decrease of $\kappa_{xx}^{\rm ph}$ 
by the extra scattering effect of phonons by magnetic skyrmions 
as suggested in the case of ferro-SkL in GaV$_4$Se$_8$ \cite{Akazawa2022}. 
\par
In our experiment at $T=3$--8\,K, there is a negative magnetothermal conductivity observed up to 8\,T (Fig.~\ref{fig3}(a)), 
which should be of magnon origin. 
It is known that a negative magnetothermal conductivity of $\kappa_{xx}^{\rm ph}$ 
is caused by a resonance scattering of phonons with spins\cite{Berman1976}. 
In that case, its field dependence scales with $B/T$, taking the minimum value when the Zeeman energy matches 
the thermal energy scale, $B \sim 4k_B T$, at which the phonon distribution reaches the maximum. 
However, the field dependence of $\kappa_{xx}$ clearly does not scale with $B/T$ 
(Fig.~\ref{fSkxx} in Appendix). 
Therefore, we conclude that a magnetic excitation possibly related to the spiral spin liquid phase\cite{Gao2017} 
would be the origin of the negative magneto thermal conductivity. 
We further mention that magnons are good quasi-particles known to have a long lifetime 
at low temperature\cite{Streib2019,Zhitomirsky2013,Chernyshev2016}, 
and they suffer resonance scattering with phonons only at high energies where the magnon branches cross the phonon ones. 
The positive magneto thermal conductivity at high fields 
is due to the enhancement of $\kappa_{xx}^{\rm ph}$ caused by the suppression of magnetic fluctuations by the magnetic field. 
\par
We finally note that phonons are unlikely the origin of the thermal Hall effect, 
although $\kappa_{xx}^{\rm ph}$ is substantial in the AFM-SkL phase. 
For a thermal Hall effect of phonons, the temperature dependence of $\kappa_{xy}$ 
is known to scale with that of $\kappa_{xx}$ as observed in several materials \cite{Li2020, Grissonnanche2020, Akazawa2020}. 
It is clearly not the case for MnSc$_2$S$_4$ as shown in Fig.~\ref{fig3}(b); 
whereas $\kappa_{xx}/T$ monotonically decreases as lowering $T$, 
$\kappa_{xy}/T$ shows a peak at around $T_N/2$. This supports the magnon origin for 
the thermal Hall effect in the AFM-SkL phase.
%*%*%*%*%*%*%*%*%*%*%*%*%*%*%*%*%*%*%*%*%*%*%*%*%*%*%*%*%*%
%*%*%*%*%*%*%*%*%*%*%*%*%*%*%*%*%*%*%*%*%*%*%*%*%*%*%*%*%*%
%*%*%*%*%*%*%*%*%*%*%*%*%*%*%*%*%*%*%*%*%*%*%*%*%*%*%*%*%*
%
%
%*%*%*%*%*%*%*%*%*%*%*%*%*%*%*%*%*%*%*%*%*%*%*%*%*%*%*%*%*%*%*%*%*%*%*%*%*%*%*%*%*%*%*%*%*
%*%*%*%*%*%*%*%*%*%*%*%*%*%*%*%*%*%*%*%*%*%*%*%*%*%*%*%*%*%*%*%*%*%*%*%*%*%*%*%*%*%*%*%*%*
%*%*%*%*%*%*%*%*%*%*%*%*%*%*%*%*%*%*%*%*%*%*%*%*%*%*%*%*%*%
\begin{figure*}[tbp]
   \centering
   \includegraphics[width=17cm]{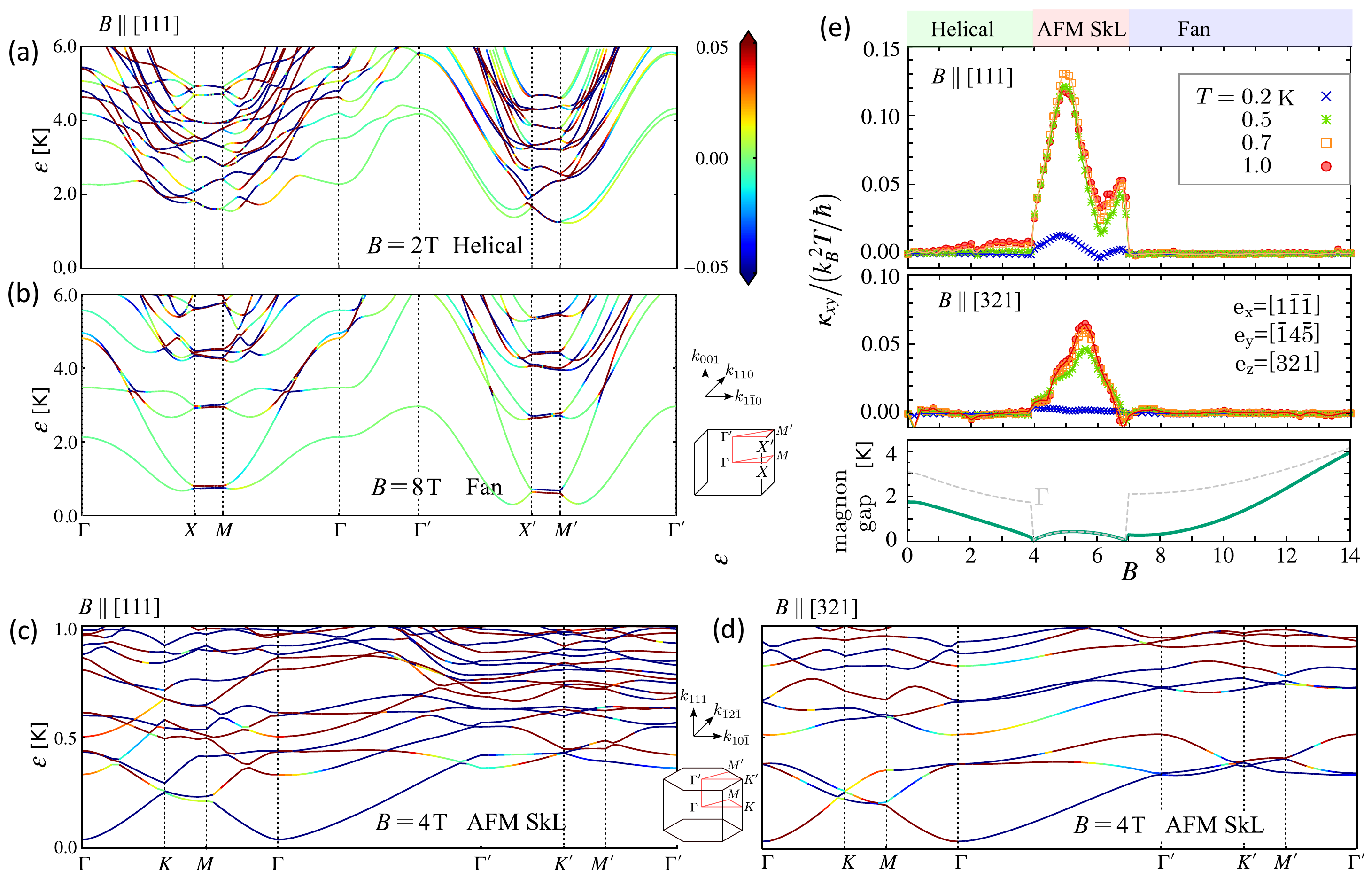}
   \caption{
   (a-d) Magnon bands obtained by the spin-wave theory for 
   the helical phase at $B=2$\,T ($N_c=16$ site unit cell, $\bm B\parallel [111]$), 
   the fan phase at $B=8$\,T ($N_c=16$, $\bm B\parallel [111]$), 
   and the AFM-SkL phase at $B=4$\,T ($N_c=384$, $\bm B\parallel [111]$ and $[321]$). 
   Reciprocal space is shown in the inset. 
   Energy bands are shown with the color density plot of the Berry curvature $\Omega^{(n)}_{xy}$ 
   at each $\bm k$-point. 
   (e) Field dependent $\kappa_{xy}/T$ for $T=0.2, 0.5, 0.7,1$\,K 
   from the linear-spin-wave theory using the same condition as panels (a)--(d). 
   Field direction is taken as $\bm B\parallel$ [111] and $[321]$. 
   The bottom panel shows the magnon gap (solid line) and the gap at $\Gamma$-point (broken line). 
}
\label{fig2}
\end{figure*}
%*%*%*%*%*%*%*%*%*%*%*%*%*%*%*%*%*%*%*%*%*%*%*%*%*%*%*%*%*%
%*%*%*%*%*%*%*%*%*%*%*%*%*%*%*%*%*%*%*%*%*%*%*%*%*%*%*%*%*%
%%%%%%%%
%%%%%%%%
\section{Theoretical}
\subsection{model}
We adopt the microscopic lattice model proposed in earlier work, 
that reproduces well the experimental observations in MnSc$_2$S$_4$ in 
their Monte Carlo simulation\cite{Gao2020}. 
The Hamiltonian is given as 
\begin{eqnarray}
{\mathcal H}&=& \sum_{\bm r, {\bm \delta_l}}\frac{J_l}{2}\bm S_{\bm r} \cdot \bm S_{\bm r+ {\bm \delta_l}} 
+ \frac{3}{2} J_\parallel \sum_{\bm r, \bm \delta_1}
 (\bm S_{\bm r}\cdot \hat{\bm \delta_1}) (\bm S_{\bm r+ {\bm \delta_l}} \cdot \hat {\bm \delta_1} )
\nonumber \\
&& +A_4 \sum_{\bm r,\mu=x,y,z} (S_{\bm r}^\mu)^4 - g \mu_B \sum_{\bm r}  \bm B \cdot \bm S_{\bm r}. 
\label{eq:ham}
\end{eqnarray}
Here, $S=5/2$ and$(J_1,J_2,J_3)=(-0.31,0.46,0.087)$\,K are the Heisenberg exchange interactions and 
${\bm \delta_l}\,(l=1,2,3)$ are 
the corresponding vectors representing the first, second, and third neighbors in the diamond lattice 
($\hat{{\bm \delta_l}}$ is the unit vector). 
The anisotropic coupling constants are set to $J_\parallel=0.01$\,K and $A_4=0.0016$\,K (in unit of temperature)\cite{Gao2020}. 
The details of the lattices parameters are shown in Appendix~\S.\ref{app:th-sw}. 
\par
We adopted the magnetic structures of the helical, fan, and AFM-SkL phases 
given in Appendix Eqs.(\ref{eq:mag1},\ref{eq:mag2}). 
The magnetic periods are fixed whereas the amplitude of uniform magnetic moment $M$ 
is determined to minimize the summation of 
classical ($E_{cl}$) and quantum zero-point fluctuation energy ($E_{qc}$). 

%*%*%*%*%*%*%*%*%*%*%*%*%*
\subsection{Large-scale spin wave theory}
Unlike the standard insulating ferro or antiferromagnets, 
whether and when the thermal Hall conductivity becomes finite in SkL is not well understood. 
The spin-wave theory for N\'eel type (ferro)-SkL on a triangular lattice 
is performed for a model with Heisenberg and DM interactions and single-ion uniaxial anisotropy 
\cite{Roldan-Molina2016, Diaz2020, Garst2017}, while they did not consider the transport properties. 
The thermal Hall effect observed in the ferromagnetic N\'eel SkL in GaV$_4$Se$_8$ 
is studied by the phenomenological U(1) gauge theory, 
showing a good agreement with the experimental data\cite{Akazawa2022}. 
However, their Chern number and the Berry curvature contradict those of the spin-wave theory. 
Indeed, the Berry curvature depends much on the details of the Hamiltonian 
and the inter-band transition, and accordingly, the same SkL structure does not 
necessarily yield the same Berry curvature. 
In such a case, the simplest U(1) gauge theory may not be sufficient. 
\par
Therefore, taking advantages about the knowledge of Eq.(\ref{eq:ham}) 
we performed a spin-wave theory for parameters where the magnetic orderings of large 
spatial periods take place, given as Eqs.(\ref{eq:mag1}) and (\ref{eq:mag2}). 
We perform a local gauge transformation to rotate the local spins to $z$-direction, 
and apply a Holstein-Primakoff transformation\cite{holstein1940} in the rotating frame, 
solving the resultant large-scale spin-wave Hamiltonian 
represented by the $2N_s\times 2N_s$ matrix, 
where we take $N_s=16$ for the helical phase and $N_s=384$ for the AFM-SkL phase\cite{colpa1978}. 
For more details of the results, see Appendix \S.\ref{app:th-magband}. 
\par 
The magnon dispersions in an applied field 
are shown in Figs.~\ref{fig2}(a) and \ref{fig2}(b) for the helical and fan phases 
and for AFM-SkL phases in Figs.~\ref{fig2}(c) and \ref{fig2}(d) for two different field directions. 
The $n$-th magnon bands are colored by the density of Berry curvature $\Omega_{xy}^{(n)}$ they carry. 
The thermal Hall conductivity is evaluated by integrating $\Omega_{xy}^{(n)}$ as\cite{matsumoto2014}, 
\begin{equation}
\kappa_{xy}=-\frac{k_{B}^{2}T}{\hbar}\int_{\mathrm{BZ}}\frac{d^{3}\bm{k}}{(2\pi)^{3}}
\sum_{n=1}^{N_s} \:c_{2}[f(\varepsilon_{n}(\bm{k}))] \:\Omega_{xy}^{(n)}(\bm{k}), 
\label{eq:Thermal-Hall}
\end{equation}
where $f(\varepsilon)=1/\{\mathrm{exp}(\varepsilon/k_{B}T)-1\}$ is the Bose distribution function, 
and $c_{2}[x]=\int_{0}^{x}dt[\mathrm{ln}\{(1+t)/t\}]^{2}$. 
\par
The calculated field-dependence of $\kappa_{xy}$ is shown in Fig.~\ref{fig2}(e). 
It remains small in the helical phase, 
reflecting the observation in Fig.~\ref{fig2}(a) that $\Omega_{xy}^{(n)}$ 
has finite contributions on part of the energy bands, 
while they are both positive and negative on nearby branches which mostly cancel out. 
If we perform the calculation by setting $J_\parallel=0$, this small contribution disappears 
and $\kappa_{xy}=0$ is obtained. 
This is because the anisotropic bond-dependent exchange interactions 
are the source of the Berry curvature\cite{Mcclarty2018}. 
\par
Contrastingly, for AFM-SkL phase, $\Omega_{xy}^{(n)}$ takes 
overall large values throughout the whole energy bands, 
particularly, a large negative value on the lowest branch when $\bm B\parallel$[111]. 
This explains why $\kappa_{xy}$ abruptly increases in entering the AFM-SkL phase. 
In further increasing the field, $\kappa_{xy}$ show peaks at 5 and 7\,T for $\bm B\parallel$[111] 
and at around 5.5\,T for $\bm B\parallel$[321]. 
The field $\bm B\parallel$[111] is perpendicular to the SkL plane and has a larger effect than $\bm B\parallel$[321]; 
the energy bands at $\varepsilon \gtrsim 0.5$\,K are much dense and $\kappa_{xy}$ are larger. 
\par
In the fan phase above 7\,T, $\kappa_{xy}$ almost dissapears, 
since $\Omega_{xy}^{(n)}$ (see Fig.~\ref{fig2}(b)) is mostly small except near the zone boundary. 
This does not explain a finite, almost field-independent $\kappa_{xy}$ in the experiment. 
It should be noted that the energy scale of $B\gtrsim $ 8\,T is comparable to or higher 
than the temperature at which the classical spiral-spin liquid phase appears at zero-field 
characterized by the coexistent magnetic correlations of several different periods\cite{Gao2017}.  
This competition may transform the system into slightly different types of orderings or another liquid state. 
Namely, the state may not be the fan phase, which is beyond the description of Eq.(\ref{eq:ham}). 
Apart from this issue, the consistency of theory and experiment at $B\lesssim 8$T is sufficient to 
confirm that AFM-SkL phase yields substantial and stable $\kappa_{xy}>0$. 
%%%%%%%%%%%%%%%
%%%%%%%%%%%%%%%
\subsection{SU(3) gauge theory}
\begin{figure*}[tbp]
   \centering
   \includegraphics[width=17cm]{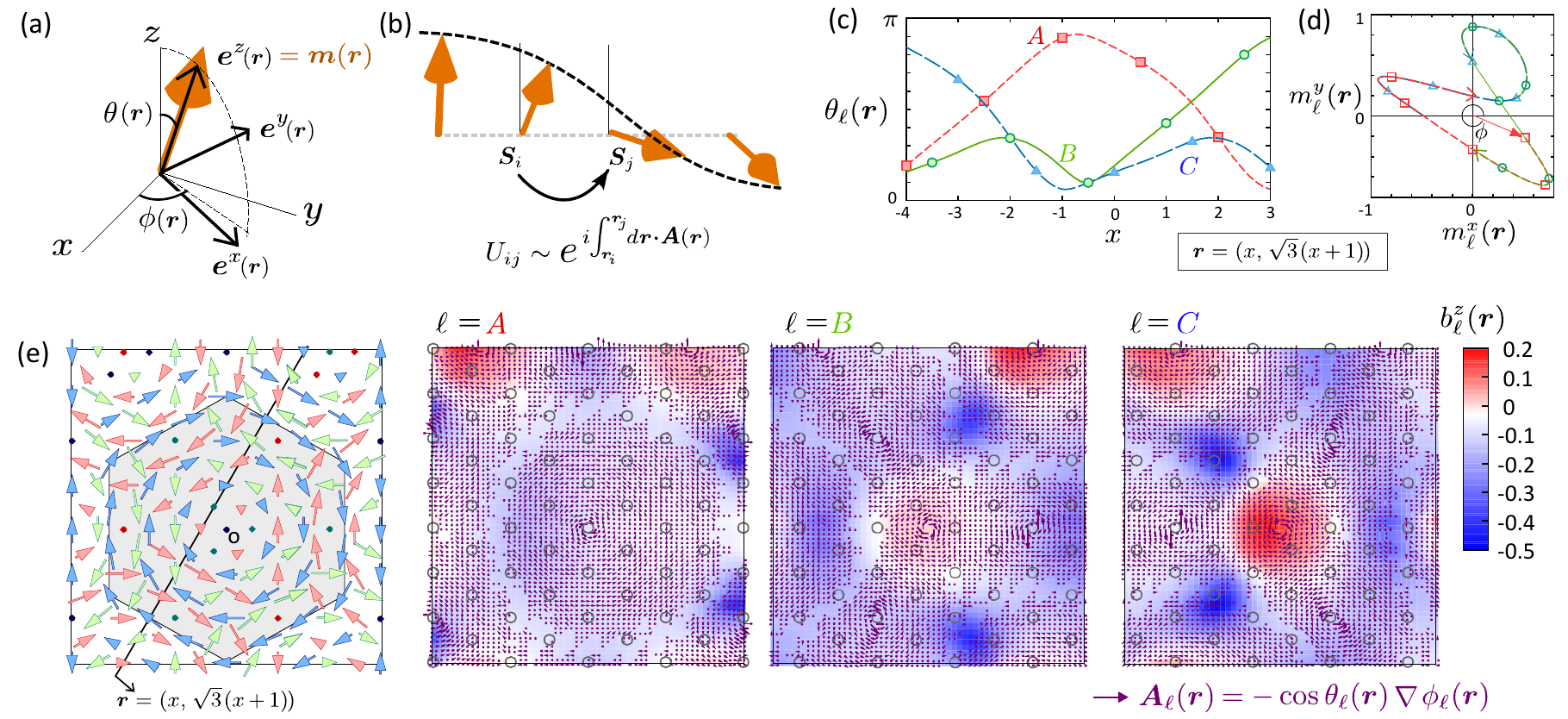}
   \caption{
(a,b) Schematic description of the ordered spin moments slowly varying in space. 
The $x,y,z$-component of spin $\bm S(\bm r)$ point toward $(\bm e^x(\bm r),\bm e^y(\bm r),\bm e^z(\bm r))$ 
which form the local coordinate with $\bm e^z(\bm r)=\bm m(\bm r)$ specified by the angle $(\theta(\bm r),\phi(\bm r))$. 
(c,d) $\theta_\ell(\bm r)$ and $m_\ell^x(\bm r)$ versus $m_\ell^y(\bm r)$ along the line, 
$\bm r=(x,\sqrt{3}(x+1))$, shown in panel {\bf d}, whose tangent give the angle $\phi_\ell$. 
(e) The right three panels show the detailed information of the U(1) gauge field based on the 
independent three sublattices; $\bm A_\ell(\bm r)=-\cos\theta_\ell(\bm r)\nabla\phi_\ell(\bm r)$ as allows 
and $b_\ell^z(\bm r)$ as density plot. Small circles indicate the location of sublattice sites. 
Left panel is the corresponding AFM-SkL structure on the triangular lattice colored red, 
green and blue for $\ell=A,B,C$ sublattices, respectively. 
}
\label{fig4}
\end{figure*}

As mentioned earlier, a field-theoretical approach does not provide us with the 
quantitative understanding on the wave functions and related physical properties. 
Still, it may give a qualitative and intuitive understanding on how 
the magnons acquire their topological nature. 
In ferromagnets, the effect of DM interactions or the non-coplanar magnetic orderings 
were straightforwardly described by the U(1) gauge that served as a vector potential for magnons. 
In a similar context, the effect of magnetic texture of ferro-SkL 
is approximated as the U(1) gauge\cite{Hoogdalem2013,Kim2019}. 
Let us give an overview; 
the effect of slowly varying magnetic moments of ferro-SkL is naturally expressed 
by a single field operator $\hat {\bm s}(\bm r)$, 
and the standard Heisenberg Hamiltonian on a discrete regular triangular lattice 
with lattice spacing $a$ is converted 
by taking the continuum limit, 
$\hat{\bm  S}_i\rightarrow \nu \hat{\bm s}(\bm r)$ with unit cell volume $\nu=\sqrt{3}a^2/2$; 
\begin{equation}
{\mathcal H}_{\rm eff}^{\rm FM} 
\sim 3J \nu \int d^2\bm r  \hat{\bm s}^t(\bm r) \big(1+\frac{a^2}{4}\nabla^2 \big) \hat{\bm s}(\bm r). 
\label{eq:ferrosk}
\end{equation}
The Holstein-Primakoff transformation to the field operator is given as 
\begin{equation}
 \hat{\bm s}(\bm r) \sim \sqrt{\frac{S}{2\nu}}\big((b_{\bm r}+b_{\bm r}^\dagger)\bm e^x(\bm r)
-i (b_{\bm r}-b_{\bm r}^\dagger)\bm e^y(\bm r)\big)
+ (\frac{S}{\nu} - b_{\bm r}^\dagger b_{\bm r}) \bm e^z (\bm r), 
\label{eq:holstein}
\end{equation}
where the unit vectors $\bm e^{\mu}(\bm r) (\mu=x,y,z)$ 
form a {\it local} orthogonal coordinate, 
with $\bm e^{z}(\bm r)\equiv \bm m(\bm r)$ pointing 
in the direction of the ordered local moment 
specified by the angles $(\theta(\bm r),\phi(\bm r))$ (see Fig.~\ref{fig4}(a)). 
Substituting Eq.(\ref{eq:holstein}) to Eq.(\ref{eq:ferrosk}) and by keeping derivative of $\bm e^{\mu}$ 
up to the first order, we obtain 
\begin{equation}
 {\mathcal H}_{\rm eff}^{\rm FM} \sim 6J\nu \int d^2\bm r\; b^\dagger_{\bm r} \big(\nabla - i \bm A(\bm r))^2 b_{\bm r} \:, 
\label{eq:ferrofinal}
\end{equation}
where $\bm A(\bm r)=-\cos\theta(\bm r) \nabla \phi(\bm r)$ serves as a fictitious U(1) vector potential 
and is generated by the spatial variation of angles $\theta$ and $\phi$ (see Fig.~\ref{fig4}(b)). 
When we put it back to the bosons $b_i$ on lattice sites, we obtain 
${\mathcal H}_{\rm eff}^{\rm FM} \sim JS \sum_{i,j} U_{ij}b^\dagger_i b_j$ 
with the U(1) gauge given as $U_{ij}={\rm exp}\big({i\int_{\bm r_i}^{\bm r_j} d\bm r \cdot\bm A(\bm r)}\big)$. 
\par 
In the present AFM-SkL, three sublattices equivalently form a large-scale SkL structure. 
This requires at least three field operators $\hat{\bm s}_\ell(\bm r) (\ell=A,B,C)$ 
that describe the spatial variation 
of moments of the three sublattices (Figs.~\ref{fig4}(c),(d)). 
The Heisenberg Hamiltonian is given by the coupling between them as 
\begin{equation}
 {\mathcal H}_{\rm eff}^{\rm AFM}\!\sim\! \frac{J \nu}{2} \int d^2\bm r \sum_{\ell,\ell'}
\hat{\bm s}_\ell^t(\bm r) \big(1+\frac{a^2}{4}\nabla^2\big) \hat{\bm s}_{\ell'}(\bm r). 
\label{eq:afsk}
\end{equation}
Here, we discard the magnetic anisotropy terms of Eq.(\ref{eq:ham}) which does not deteriorate 
the following discussions. 
Introducing three species of Holstein-Primakoff bosons $b_\ell(\bm r)$ 
by adding indices $\ell=A,B,C$ to Eq.(\ref{eq:holstein}), 
we reach the following effective Hamiltonian for the AFM-SkL, 
\begin{eqnarray}
{\mathcal H}_{\rm eff}^{\rm AFM}\!\sim\!\int \! d^{2}\bm{r} 
{\Phi}^{\dagger}(\bm{r})\big[ 
\bm F(\bm{r})a^{2}\nabla^{2}\!+\!\sum_{\mu=x,y}\!\bm G_{\mu}(\bm{r})a\partial_{\mu}
\big] {\Phi}(\bm{r}),
\end{eqnarray}
where $\Phi^\dagger(\bm r)=(b_A^\dagger, b_A, b_B^\dagger, b_B, b_C^\dagger, b_C)$ 
and $\bm F(\bm r)$ and $\bm G_{\mu}(\bm r)$ are $6\times 6$ matrices 
consisting of $\bm e^\mu_\ell$ and $\bm e^\mu_{\ell'}$ 
and $\bm e^\mu_\ell\cdot \partial_\mu \bm e^\mu_{\ell'}$, respectively, with $\ell\ne \ell'$. 
In particular, the diagonal elements of $\bm G_{\mu}(\bm{r})$ 
are proportional to the U(1) gauge fields of the three sublattices, 
$\bm A_\ell(\bm r)=-\cos\theta_\ell(\bm r)\partial_\mu \phi_\ell(\bm r)$. 
We show in Fig.~\ref{fig4}(e) the profiles of $\bm A_\ell(\bm r)$ 
calculated as continuous two-dimensional vector fields and the 
corresponding fictitious fields are given as 
\begin{equation}
b^z_\ell(\bm r)= \partial_x A_\ell^y(\bm r)-\partial_y A_\ell^x(\bm r)
\end{equation}
(see Appendix \S.\ref{app:th-u1} for details).
We see that the locations where $\bm m(\bm r)$ points in the $z$-direction ($\theta=0,\pi$) 
serve as vortex centers of $\bm A_\ell(\bm r)$ and generate large $b_\ell^z(\rm r)$. 
\par
However, there are substantial off-diagonal elements in $\bm G_\mu(\bm r)$ that 
couple these three U(1) gauge fields and transform them to the SU(3) gauge fields. 
Let $\Psi^\dagger (\bm r)$
be the operators obtained by the unitary transformation to $\Phi^\dagger(\bm r)$ that diagonalizes $\bm F(\bm r)$, 
where we find the form 
\begin{equation}
\hat H_{\rm eff}^{\rm AFM} \!\sim\! \int d^2\bm r \Psi^\dagger (\bm r) 
\big[\sum_{\mu=x,y}\left(I_{3\times3}\partial_{\mu}+\bm T_{\mu}(\bm{r})\right)^{2}\otimes\tau^{x}\big]\Psi (\bm r), 
\end{equation}
where $\bm G_\mu(\bm r)$ is converted to the  $3\times 3$ matrix $\bm T_\mu(\bm r)$ 
that serves as a high-rank vector potential for the SU(3) gauge field, 
as we find from the analogy with Eq.(\ref{eq:ferrofinal}). 
%*%*%*%*%*%*%*%*%*%*%*%*%*%*%*%*%*%*%*%*%*%*%*%*%*%*%*%*%*%
%*%*%*%*%*%*%*%*%*%*%*%*%*%*%*%*%*%*%*%*
%*%*%*%*%*%*%*%*%*%*%*%*%*%*%*%*%*%*%*%*%*%*%*%*%*%*%*%*%*%
%*%*%*%*%*%*%*%*%*%*%*%*%*%*%*%*%*%*%*%*%*%*%*%*%*%*%*%*%*
%*%*%*%*%*%*%*%*%*%*%*%*%*%*%*%*%*%*%*%*
\section{ Summary and Discussion }
We have experimentally measured the thermal conductivity $\kappa_{xx}$ and 
thermal Hall conductivity $\kappa_{xy}/T$ of the insulating magnet MnSc$_2$S$_4$ 
in an applied field up to 14\,T and found a substantial thermal Hall effect 
that emerges in the AFM-SkL phase in an applied field of $B\sim$ 4--8\,T at temperatures 
$T\lesssim $1\,K lower than $T_N$. 
By performing a large-scale spin wave theory considering the Hamiltonian specific to the material, 
we showed that the AFM-SkL exhibits a very dense magnon band structure, 
and hosting a large amount of Berry curvature. 
The corresponding thermal Hall coefficient calculated from Helical to AFM-SkL phases in increasing the magnetic field 
consistently explains the experimental observations. 
\par 
We now discuss in more detail the correspondence of the experiment in Fig.~\ref{fig1}(e) and the theory 
in Fig.~\ref{fig2}(e) for $\bm B\parallel$[321]. 
Although we cannot precisely refer to the small structures in the experimental data 
that may depend on sample qualities, the common features are captured very well; 
at 0.2\,K, $\kappa_{xy}$ remains small and structureless, while for 0.5, 0.7, and 1\,K, there 
is a single peak at around 5.5--6\,T and the amplitudes no longer differ much for different temperatures. 
To understand these features, 
we calculate how many magnon bands from the bottom would give major contributions to $\kappa_{xy}$ 
(Appendix Fig.~\ref{fStheory2}). 
At 0.2\,K, including 24 bands already reproduces well the results of the full 384 bands, 
while at 0.5\,K, the 24 bands corresponding to the energy window of up to 2\,K are not at all sufficient. 
This means that the dense band structure of skyrmions including those at high energies plays a crucial role 
in magnon transport. 
Therefore, although the magnon gap increases toward 5\,T and then decreases with a field, 
$\kappa_{xy}$ is not sensitive to this gap opening. The nontrivial distribution of $\Omega_{xy}^{(n)}$ up to the high energy 
determines the field dependence of $\kappa_{xy}$. 
\par
In this context, one needs to be careful about 
the Chern numbers or equivalently the Berry curvatures that depend on the details of the Hamiltonian. 
Even for the simple ferromagnetic N\'eel SkL phase of GaV$_4$Se$_8$ \cite{Akazawa2022}, 
although the phenomenological U(1) gauge theory give $\kappa_{xy}$ in good agreement with the experimental data, 
the corresponding Chern number contradicts those obtained by the spin-wave theory\cite{Diaz2020,Molina2016} 
at the lowest two energy bands. 
The values of $\kappa_{xy}$ that have contributions of Berry curvature from a wide range of bands 
are particularly sensitive to the details 
and need an accurate treatment for their quantitative evaluation. 
\par 
However, an overall origin of the emergent Berry curvature of magnon bands in the AFM-SkL is the 
SU(3) gauge as we discussed earlier. 
To give an intuitive understanding, it is useful to refer to the antiferromagnets on bipartite lattices, 
where the two species of magnons on the two sublattices can be regarded as up and down pseudo-spin ones. 
When the antisymmetric or spatially anisotropic exchange couplings are introduced, 
they serve as ``pseudo-spin-orbit coupling", which mixes the motion of up and down magnons
\cite{kawano2019-1,kawano2019-2,kawano2019-3}. 
This effect is represented by the SU(2) gauge field on magnons in analogy with the SU(2) hopping of 
Rashba electrons of semiconductors\cite{rashba1960}, 
and allowed the thermal Hall effect on a square lattice antiferromagnet\cite{kawano2019-2} 
which was prohibited in the U(1) picture\cite{katsura2010}. 
\par
In the present triangular-based system, we have three types of magnons living on the three sublattices, 
which can be regarded as three-component pseudo-spins. 
If the magnetic moments form a regular coplanar 120$^\circ$ ordering, we find $\bm G_\mu(\bm r)=0$, namely 
we do not have the spatial variation of $\hat s_\ell (\bm r)$ for the U(1) gauge field. 
In the AFM-SkL phase, the U(1) gauge fields are switched on, and at the same time 
they communicate with each other through the ``pseudo-spin-orbit coupling" given by 
the off-diagonal $\bm G_\mu(\bm r)$ that originate
from the exchange interactions between spins on different sublattices. 
The SU(3) pseudo-spin represents the sublattice degrees of freedom 
and the "orbit" refers to the kinetic motions of magnons. 
\par
The SU(3) gauge has been a theoretical object for describing the interactions of 
quarks or gluons in particle physics, and not much has been discussed in condensed matter, particularly 
in experiment. 
While the SU(3) symmetry can sometimes appear, e.g. in cold atoms \cite{Molina2009} 
and the SU(3) Heisenberg Hamiltonian may yield exotic phases\cite{Yamamoto2020}, 
the present system would be 
the first to observe experimentally the phenomena that directly reflect the SU(3) gauges in material solids. 
\\
%*%*%*%*%*%*%*%*%*%*%*%*%*%*%*%*%*%*%*%*%*%*%*%*%*%*%*
%*%*%*%*%*%*%*%*%*%*%*%*%*%*%*%*%*%*%*%*%*%*%*%*%*%*%*
\vspace{8mm}
\\

%%%%%%%%%%%%%%%%%%%%%%%%%%%%%%%%%%%%%%%%%%%%%%%%
%%%%%%%%%%%%%%% Acknowledgements %%%%%%%%%%%%%%%
%%%%%%%%%%%%%%%%%%%%%%%%%%%%%%%%%%%%%%%%%%%%%%%%
\begin{acknowledgments}
The work is supported by JSPS KAKENHI Grants  Grants No. JP19H01848, No. JP17K05533, No.
JP18H01173, No. 20K03773, No. JP21H05191, No. JP21K03440, No. 21H01035, and No. 17H06137 
from the Ministry of Education, Science, Sports and Culture of Japan 
and the Murata Science Foundation. 
\end{acknowledgments}

%*%*%*%*%*%*%*%*%*%*%*%*%*%*%*%*%*%*%*%*%*%*%*%*%*%*%*%*%*%*%*%*%*%*%*
%*%*%*%*%*%*%*%*%*%*%*%*%*%*%*%*%*%*%*%*%*%*%*%*%*%*%*%*%*%*%*%*%*%*%*
%*%*%*%*%*%*%*%*%*%*%*%*%*%*%*%*%*%*%*%*%*%*%*%*%*%*%*%*%*%*%*%*%*%*%*
%%%%%%%%%%%%%%%%%%%%%%%%%%%%%%%%%%%%%%%%
%%%%%%%%%%%%%%% Appendix %%%%%%%%%%%%%%%
%%%%%%%%%%%%%%%%%%%%%%%%%%%%%%%%%%%%%%%%
\appendix
\section{Details of the thermal transport measurements}
\label{ap:exp1}
We show the measurement setup, the magnetic field dependence of $\kappa_{xx}$, 
the field-injection process dependence of $\kappa_{xy}$, 
and the sample dependences of both of them. 
\subsection{Measurement setup}
\begin{figure}[b]
   \centering
   \includegraphics[width=6cm]{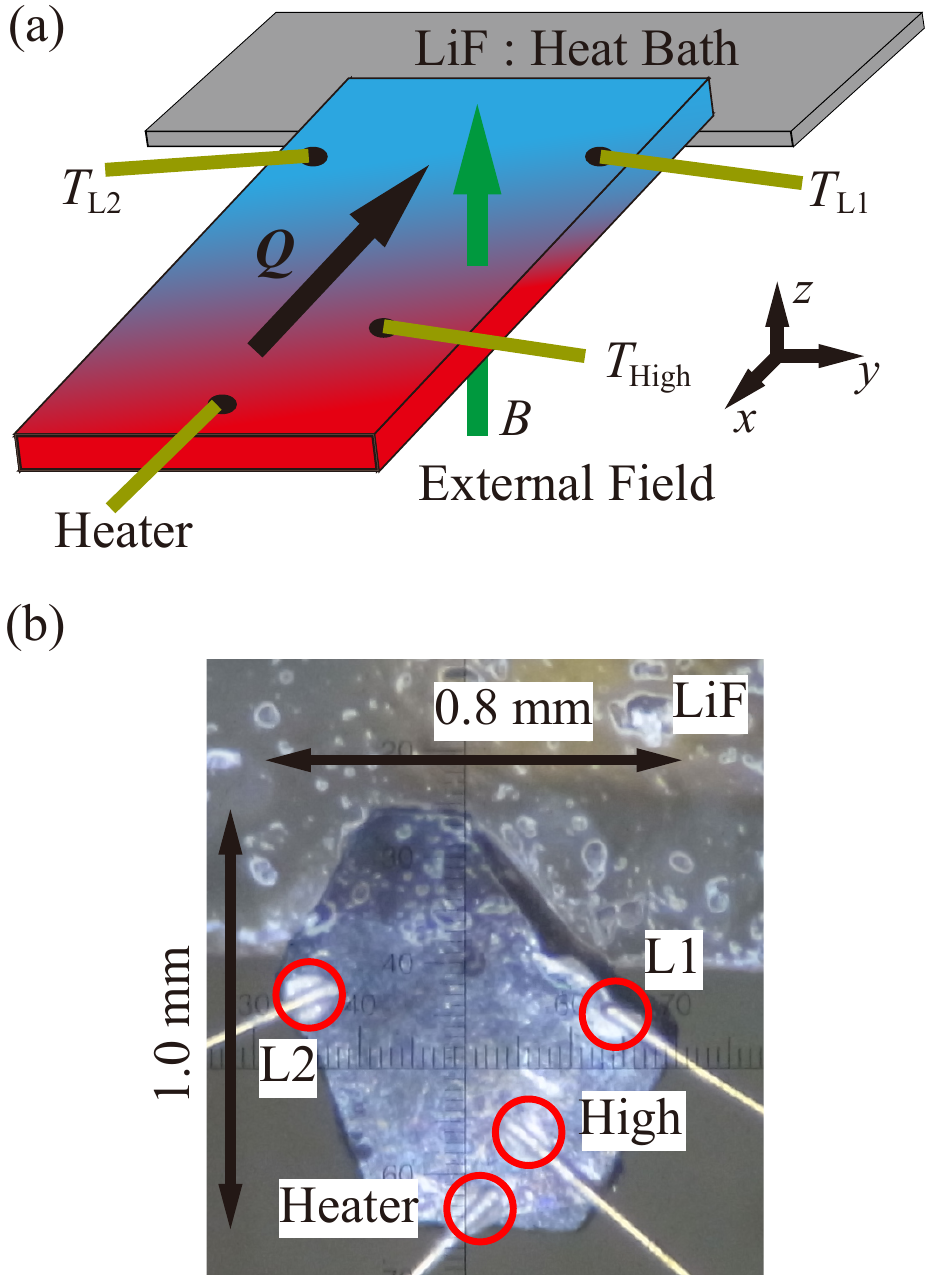}
   \caption{(a) Schematic figure of the setup of the thermal transport measurements.
(b) Photograph of the sample mounted on the heat bath of LiF. The thermal contacts for the heater and the thermometers are marked by red circles. 
}
%\vspace{-4mm}
\label{fSsamp}
\end{figure}

As shown in Figs.~\ref{fSsamp}(a) and \ref{fSsamp}(b), one heater and three thermometers 
($T_{\rm High}$, $T_{\rm L1}$, and $T_{\rm L2}$) were attached to the sample using a silver paste. 
The sample size is about $1.0\times 0.8\times 0.07$\,mm$^3$. 
To avoid a background signal coming from metal, the sample was attached to the insulating LiF heat bath with non-metallic grease. 
The heat current $J_{\rm Q}$ and the magnetic field $B$ were applied along [111] and [321] directions of the sample, respectively.
Both the longitudinal $\Delta T_x (\Delta T_x=T_{\rm High}-T_{\rm L2})$ and 
the transverse $\Delta T_y (\Delta T_y=T_{\rm L1}-T_{\rm L2})$ 
%*%*%*%*%*%*%*%*%*%*%*%*%*%*%*%*%*%*%*%*%*%*%*%*%*%*%*%*%*%
%*%*%*%*%*%*%*%*%*%*%*%*%*%*%*%*%*%*%*%*%*%*%*%*%*%*%*%*%*%
temperature differences were measured as a function of the heat current $J_Q=Q/wt$, where $Q$ is the heater power, $t$ is the thickness of the sample, and $w$ is the mean sample width. 
To cancel the longitudinal component in $\Delta T_y$ by the misalignment effect, 
$\Delta T_y$ was asymmetrized with respect to the field direction as  $\Delta T_y^{\rm asym}=\Delta T_y (+B)-\Delta T_y (-B)$. 
To take into account a magnetic hysteresis effect, 
this antisymmetrization was done separately for the data obtained in the field-up process and that in the field-down process. 
The thermal (Hall) conductivity $\kappa_{xx}(\kappa_{xy})$ is derived by 
\begin{equation}
\left(\begin{array}{c}
Q/wt \\ 0
\end{array}\right)
=
\left(\begin{array}{rr}
 \kappa_{xx} & \kappa_{xy} \\
-\kappa_{xy} & \kappa_{xx} 
\end{array}\right)
\left(\begin{array}{c}
\Delta T_x/L \\
\Delta T_y^{\rm asym}/w'
\end{array}\right),
\end{equation}
where $L$ is the length between the thermal contacts for reading $T_{\rm High}$ and $T_{\rm L1}$ 
and $w'$ is the length between the thermal contacts for $T_{\rm L1}$ and $T_{\rm L2}$. 
The irregular shape of the sample and the finite size of the thermal contacts cause 
uncertainty in estimating these geometrical factors, 
resulting in the ambiguity of the absolute values of $\kappa_{xx}$ and $\kappa_{xy}$ by a factor of 2--4. 
%*%*%*%*%*%*%*%*%*%*%*%*%*
%*%*%*%*%*%*%*%*%*%*%*%*%*

%
%*%*%*%*%*%*%*%*%*%*%*%*%*%
\begin{figure}[tbp]
   \centering
   \includegraphics[width=8.5cm]{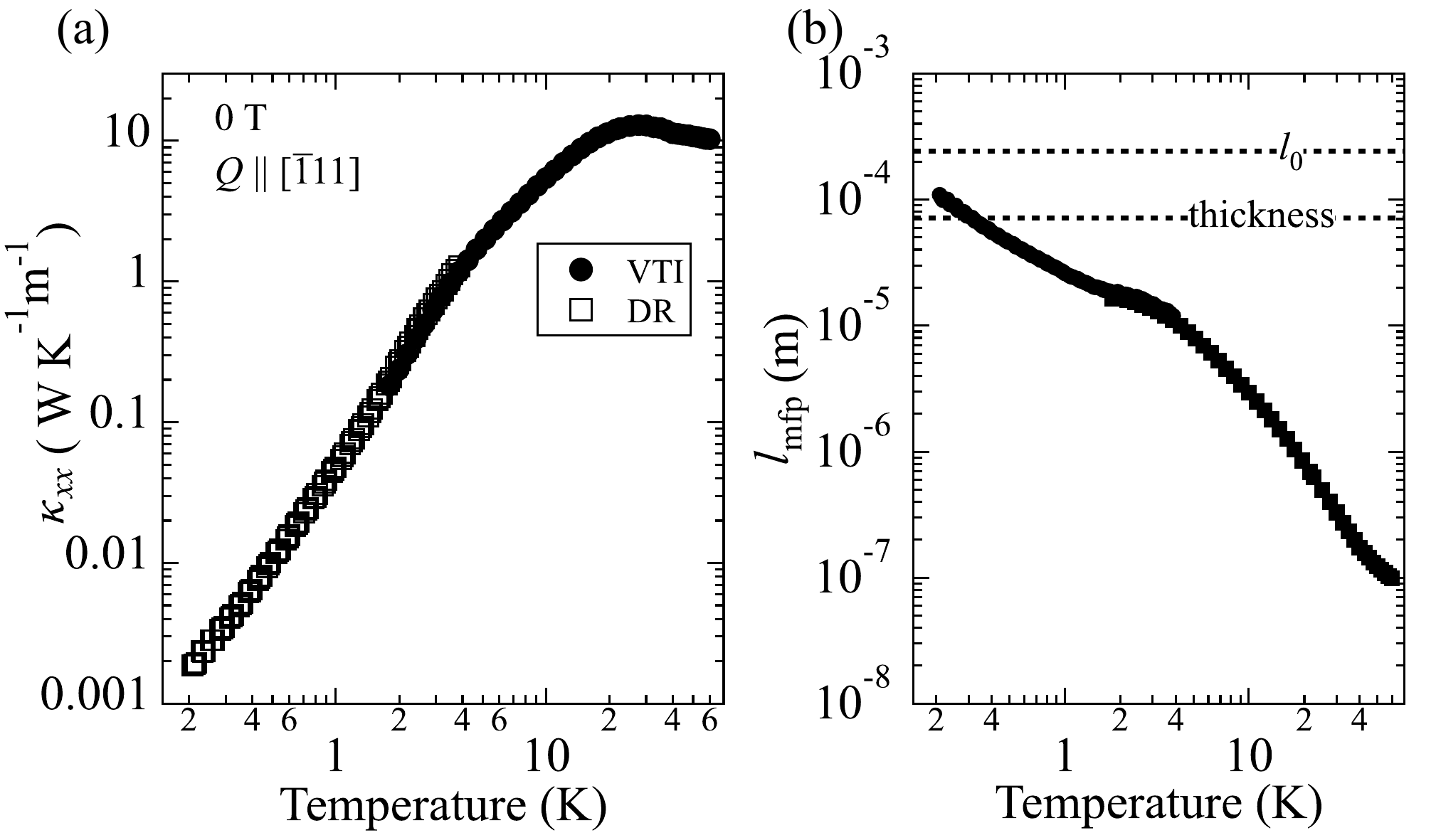}
   \caption{(a) Temperature dependence of $\kappa_{xx}$ at zero external field measured in the variable-temperature insert (VTI, circles)
and the dilution refrigerator (DR, squares). 
(b) Phonon mean free path estimated by assuming that the data in panel (a) is solely from phonons. 
}
%\vspace{-4mm}
\label{fSkxx_t}
\end{figure}
%%*%*%*%*%*%*%*%*%*%*%*%*%*
%*%*%*%*%*%*%*%*%*%*%*%*%*%
\begin{figure}[tbp]
   \centering
   \includegraphics[width=8cm]{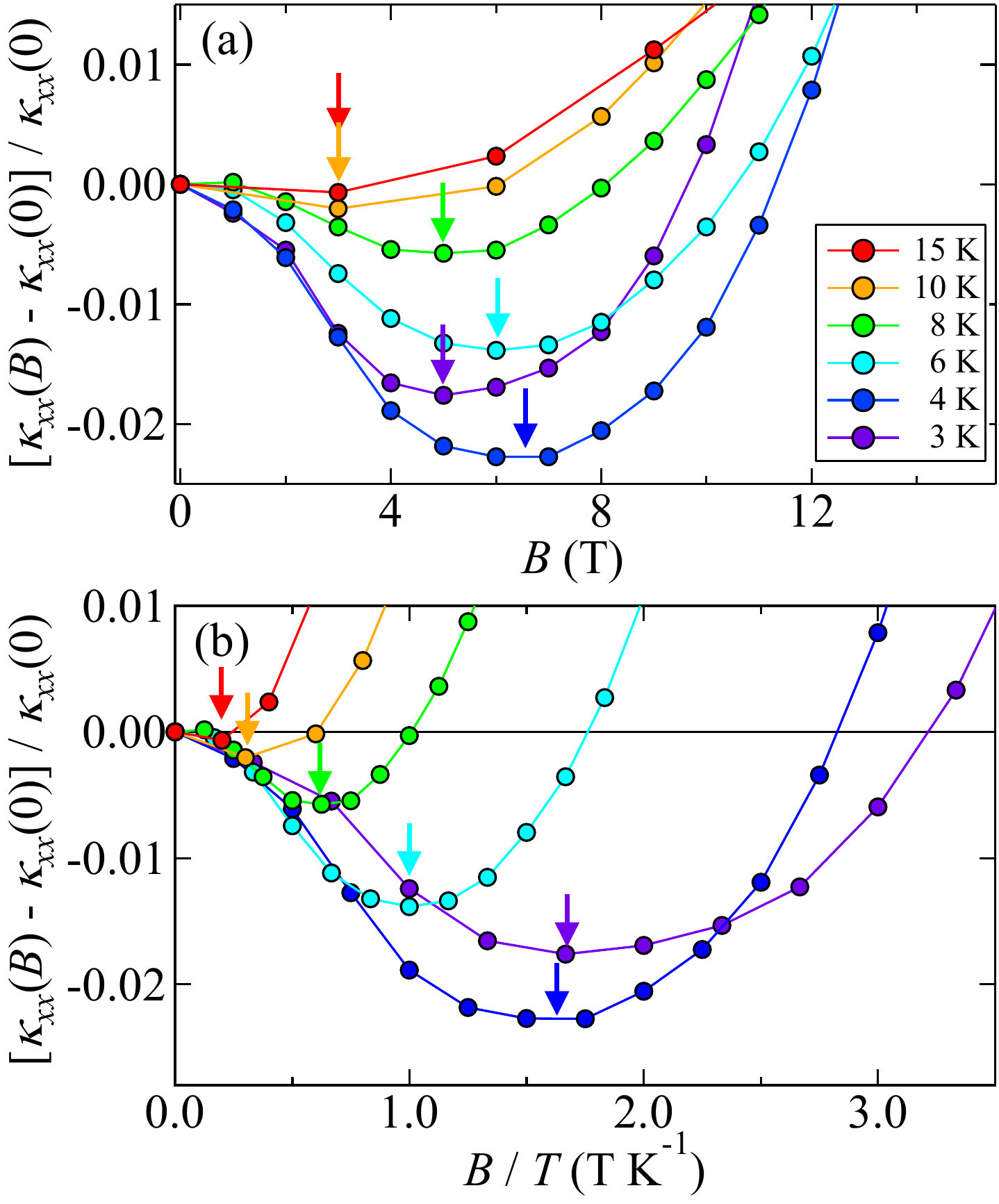}
   \caption{Normalized $\kappa_{xx}$ as a function of (a) $B$ and (b) $B/T$ for 
   several choices of temperatures above $T_N$. 
   Arrows denote the minimum position of normalized $\kappa_{xx}$.
}
%\vspace{-4mm}
\label{fSkxx}
\end{figure}
%*%*%*%*%*%*%*%*%*%*%*%*%*%
%*%*%*%*%*%*%*%*%*%*%*%*%*%
%*%*%*%*%*%*%*%*%*%*%*%*%*%*%*%*%*%*%*%*%*%*%*%*%*%*%*%*%*%
%
\subsection{Estimation of the phonon mean free path}
\label{ap:exp-mfp}
\noindent
Figure~\ref{fSkxx_t}(a) shows the temperature dependence of $\kappa_{xx}$ in a zero external magnetic field. 
It exhibits a $ T^2$ behavior without a clear anomaly at the N\'eel temperature ($T_N=2.3$\,K). 
Here, we estimate the phonon mean free path ($\ell_\textrm{mfp}$) by assuming that $\kappa_{xx}$ 
is only given by phonons ($\kappa_{xx}^\textrm{ph}$) 
as shown in Fig.~\ref{fSkxx_t}(b). 
\par
The process of evaluating $\ell_{\rm mfp}$ is given as follows. 
The thermal conductivity of phonons is given by 
\begin{equation}
\kappa_{xx}^{\rm ph}= \frac{1}{3} C_{\rm ph}v_{\rm ph}l_{\rm ph}, 
\end{equation}
where $C_{\rm ph}$, $v_{\rm ph}$, and $l_{\rm ph}$ are the heat capacity, the sound velocity, and the mean free path of phonons, respectively. 
Referring to Ref.[\onlinecite{Fritsch2004}], 
we obtain the temperature dependence of $C_{\rm ph}$ for acoustic phonons. 
The sound velocity is estimated as approximately 1770 m/s from the Debye temperature of 154 K. 

%%*%*%*%*%*%*%*%*%*%*%*%*%*
\subsection{ Magnetic field dependence of $\kappa_{xx}$}
\label{ap:exp-kxx}
\noindent
Figure~\ref{fSkxx}(a) shows $B$-dependence of $\kappa_{xx}(B)$ normalized as 
$(\kappa_{xx}(B)-\kappa_{xx}(0))/\kappa_{xx}(0)$ in the temperature range of 3\,K $\le T \le 15$\,K. 
The field at which the normalized $\kappa_{xx}(B)$ takes the minimum 
shifts to higher fields when lowering the temperature from 15\,K to 4\,K, and then slightly 
shifts to the lower field at 3\,K. 
The suppression of $\kappa_{xx}$ by the magnetic field (negative magneto thermal conductivity) 
can be attributed to 
the decrease of either $\kappa_{\rm ph}$ or $\kappa_{\rm mag}$. 
\par
The former suppression originates from the resonance scattering effect of phonons due to spins, 
and should be enhanced the most when the energy scale $4k_BT$ that gives the maximum of 
the density distribution of phonons becomes equal to the Zeeman splitting, $gS\mu_BB$. 
We plotted the normalized $\kappa_{xx}$ as a function of $B/T$ in Fig.~\ref{fSkxx}(b). 
If this scenario applies, $\kappa_{xx}$ should take the minimum at $B/T\sim 4k_B/gS\mu_B$, 
independent of $T$, which does not happen here. 
Therefore, we conclude that the suppression of $\kappa_{xx}^{\rm ph}$ by the magnetic field 
is not ascribed to the resonant scattering effect of phonons. 
\par
The remaining possibility is the decrease of $\kappa_{xx}^{\rm mag}$. 
Since the Weiss temperature is $-23$\,K, 
the antiferromagnetic correlation will develop at temperatures below 20\,K. 
In fact, the magnetic specific heat is found to increase with decreasing temperature at $T<15$\,K. 
In frustrated insulating quantum magnets, such strong spin correlation, even though they may 
not form a long-range order, is known to sometimes host a spinons excitation 
that carries heat current. 
In MnSc$_2$S$_4$, in a paramagnetic phase at low temperatures above $T_N$ is possibly a 
classical spiral-spin liquid. There, the magnetic diffuse scattering experiments 
show the existence of some sort of magnetic excitations relevant to 
a manifold of wave numbers forming surfaces in the reciprocal space, characterizing this classical spin liquid. 
Therefore, it is natural to consider that the 
external magnetic field will suppress the strong 
fluctuation of this spiral spin liquid and decreases $\kappa_{xx}^{\rm mag}$ accordingly. 
This scenario is consistent with the observed energy scales: the order of the magnetic interaction $\sim $20\,K and the location of minimum $\sim 6$\,T. 
%
%*%*%*%*%*%*%*%*%*%*%*%*%*%
\begin{figure}[tbp]
   \centering
   \includegraphics[width=7cm]{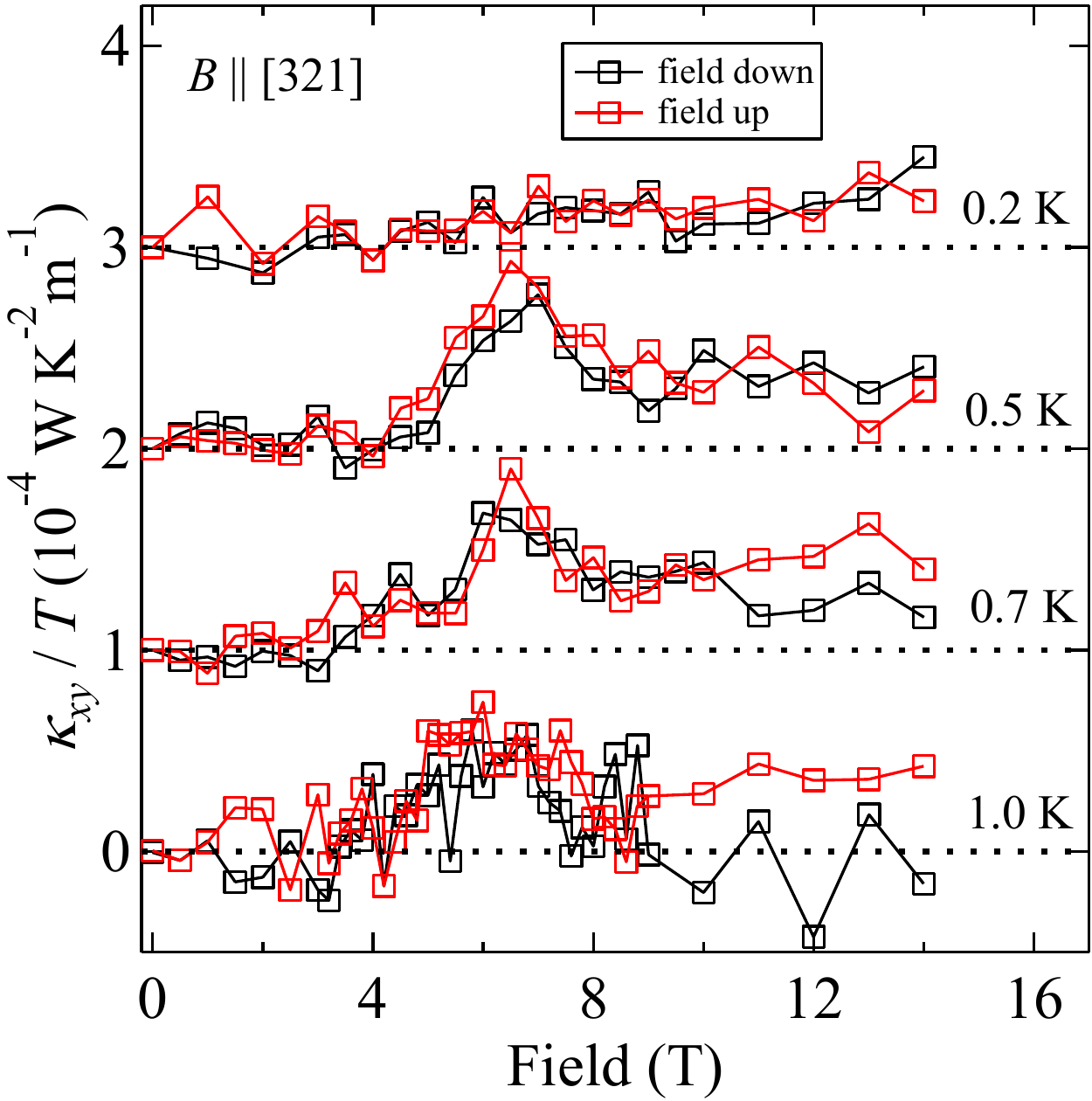}
   \caption{Magnetic field $B$ dependence of $\kappa_{xy}/T$ obtained in the field up and down processes for sample $\#$1. 
}
%\vspace{-4mm}
\label{fSkxy_ud}
\end{figure}
%*%*%*%*%*%*%*%*%*%*%*%*%*%
%*%*%*%*%*%*%*%*%*%*%*%*%*%
%*%*%*%*%*%*%*%*%*%*%*%*%*%*%*%*%*%*%*%*%*%*%*%*%*%*%*%*%*%
\subsection{ Field-up and down processes of $\kappa_{xy}/T$}
\label{ap:exp-kxy}
\noindent
Since magnetic skyrmions are a metastable magnetic order 
that can not appear without the aid of a magnetic field or strong magnetic anisotropy, 
it is often found in the previous studies
\cite{Munzer2010, Oike2016, Okamura2016, Bauer2016, Chauhan2022, Akazawa2022}
that the phase boundary (1st order phase transition) 
between SkL phase and the other phases usually vary, 
or the density of skyrmions may change according to the history of applying a magnetic field. 
The change in skyrmion density will vary the magnon density which easily influences 
the thermal transport. 
Therefore, we measured the thermal Hall conductivity for the field-up and field-down processes. 
Figure~\ref{fSkxy_ud} shows $\kappa_{xy}/T$ 
as a function of field-up and field-down for temperatures $T=0.2, 0.5, 0.7$, and 1\,K. 
We find that the two processes agree well, indicating that 
the phase boundaries do not depend on the processes. 
This behavior is in sharp contrast to the other skyrmion materials, e.g. 
GaV$_4$Se$_8$, that exhibit different behavior between field-up and down processes. 
The data points given in Fig.~\ref{fig1}(d) in the main text are the ones averaged for the two 
processes, and the error bars are the maximum deviation of the measured data from the average values. 
%
%*%*%*%*%*%*%*%*%*%*%*%*%*%
\begin{figure}[tbp]
   \centering
   \includegraphics[width=8.5cm]{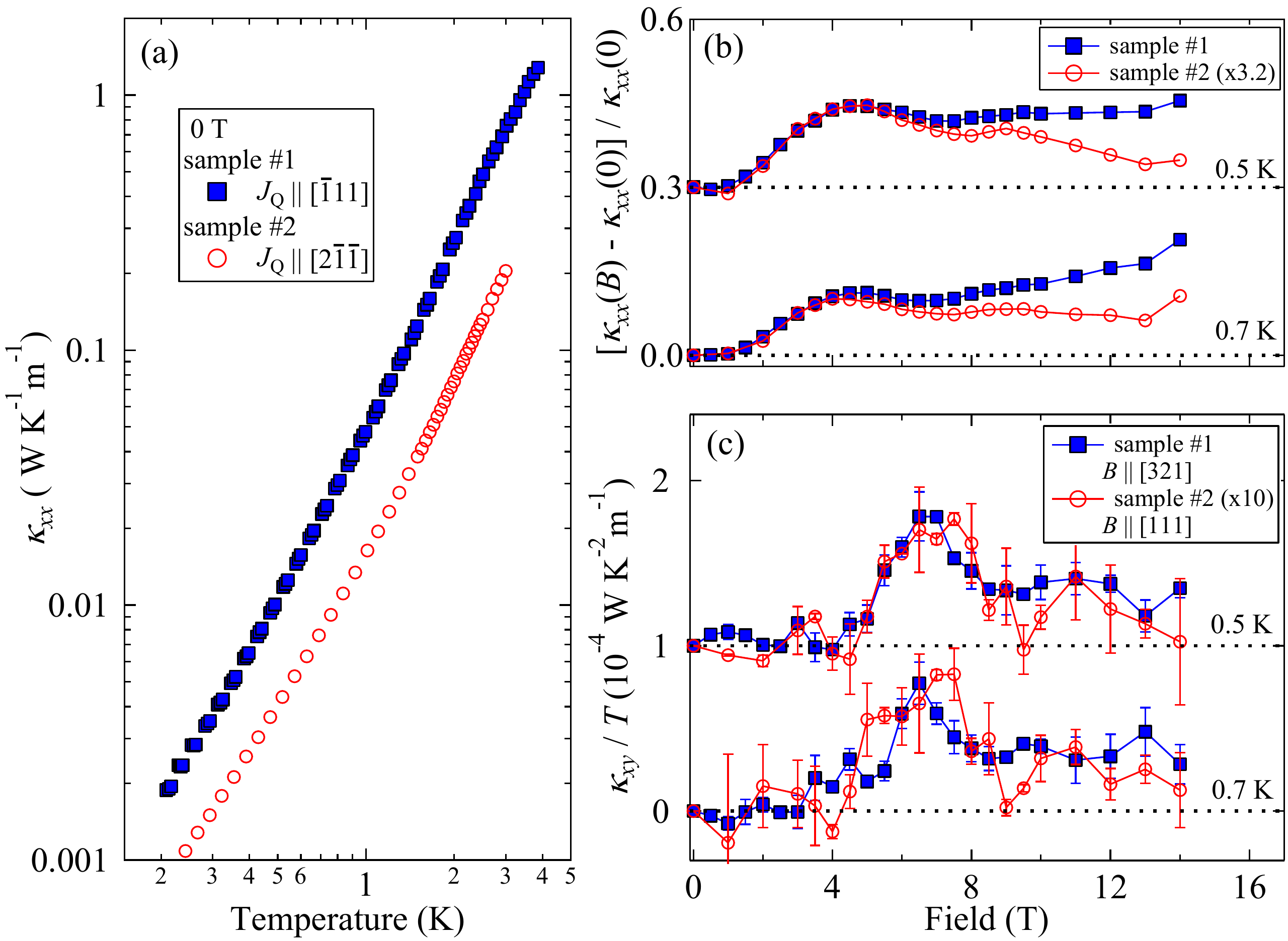}
   \caption{
(a) Temperature dependence of $\kappa_{xx}$ for sample $\#$1 and sample $\#$2. 
Field $B$ dependence of (b) normalized $\kappa_{xx}$ and (c) $\kappa_{xy}/T$ 
at $T=$0.5K and 0.7\,K obtained in the field up and down processes for 
the two samples. 
}
\label{fSsamp2}
\end{figure}
%*%*%*%*%*%*%*%*%*%*%*%*%*%
%
%*%*%*%*%*%*%*%*%*%*%*%*%*%
\subsection{Thermal transport properties of sample $\#$2}
\label{ap:exp-samp}
\noindent
The $B$-$T$ phase diagram of MnSc$_2$S$_4$ are studied in detail for 
$\bm B\parallel$ [111] and $\bm B\parallel$ [110]. 
Since for the former case, the phase boundary has a simple structure as a function of field, 
we studied sample $\#$2 which has [111] plane at the sample surface. 
Here, we compare the results of sample $\#$2 with those of sample $\#$1 shown in the main text.
\par
Figure~\ref{fSsamp2}(a) shows the temperature dependence of $\kappa_{xx}$. 
Since the sample quality is not as good for sample $\#$2 as sample $\#$1, 
the absolute values differ by a factor of two, while they show good agreement in 
their temperature dependences. 
\par
As shown in Fig.~\ref{fSsamp2}(b), the value of $\kappa_{xx}$ normalized by 
the zero-field values show very good agreement with those of sample $\#$1 for $B<9$\,T. 
The peak at around 4\,T and the dips at 7\,T are the anomalies that take place at the 
magnetic phase transition points reported previously. 
For fields higher than 9\,T, the normalized $\kappa_{xx}$ of sample $\#$2 
show strong suppression in further increasing $B$. 
This is in sharp contrast to the case of sample $\#$1. 
Since the high-field phase for $\bm B\parallel$ [111] is the multi-domain phase 
consisting of single-$Q$ state, the field angle may influence the structure of the domain 
and accordingly, the magnitude of orderings. 
Since thermal transport at low temperatures is very likely to be caused by magnons, 
the number of carriers may change due to this effect. 
\par
In Fig.~\ref{fSsamp2}(c) we show the $\kappa_{xy}/T$ as function of $B$ for 0.5\,K and 0.7\,K. 
Although the amplitude of $\kappa_{xy}/T$ of sample $\#$2 is about 1/10 of that 
of sample $\#$1, the profile of the field dependences after scaling the factor 
shows very good agreement. 
We thus expect that the magnetic phases 
show the same dependence on the field $\bm B\parallel$ [321] and $\bm B\parallel$ [111] 
below 9\,T, as has been confirmed in Figs.~\ref{fSsamp2}(a) and \ref{fSsamp2}(b). 
It has been shown previously that the AFM-SkL phase appears not only in 
the particular field direction $\bm B\parallel$ [111] but 
also along $\bm B\parallel$ [100] and $\bm B\parallel$ [110], 
although the shapes of the phase boundaries differ\cite{Rosales2022}. 
Therefore, we can safely conclude that the results we obtained for 
$\bm B\parallel$ [321] for sample $\#$1 is the one that detects the 
emergent AFM-SkL phase reported previously. 

%*%*%*%*%*%*%*%*%*%*%*%*%*%
%*%*%*%*%*%*%*%*%*%*%*%*%*%
\section{Theoretical}
\subsection{Details of the spin wave theory}
\label{app:th-sw}
We consider the spin Hamiltonian on a diamond lattice in Eq.(1) in the main text. 
The vectors 
$\bm{\hat\delta}_{1}, \bm{\hat\delta}_{2}$ and $\bm{\hat\delta}_{3}$ 
showing uniaxial spin anisotropy along the bond directions of the lattice is 
\begin{align}
4\bm{\hat\delta}_{1}
&=
\left(1,1,1\right),
\left(-1,1,1\right),
\left(1,-1,1\right),
\left(1,1,-1\right),
\nonumber \\
&\hspace{12pt}
\left(-1,-1,1\right),
\left(1,-1,-1\right),
\left(-1,1,-1\right),
\left(-1,-1,-1\right),
\nonumber \\
%%%%%%%%%%%%%%%
%%%%%%%%%%%%%%%
2\bm{\hat \delta}_{2}
&=
\left(1,0,1\right),
\left(0,1,1\right),
\left(-1,0,1\right),
\left(0,-1,1\right),
\nonumber \\
&\hspace{12pt}
\left(1,1,0\right),
\left(1,-1,0\right),
\left(-1,1,0\right),
\left(-1,-1,0\right),
\nonumber \\
&\hspace{12pt}
\left(1,0,-1\right),
\left(0,1,-1\right),
\left(-1,0,-1\right),
\left(0,-1,-1\right),
\nonumber \\
%%%%%%%%%%%%%%%
%%%%%%%%%%%%%%%
4\bm{\hat \delta}_{3}
&=
\left(3,1,1\right),
\left(-3,1,1\right),
\left(3,-1,1\right),
\left(3,1,-1
\right),
\nonumber \\
&\hspace{12pt}
\left(-3,-1,1\right),
\left(3,-1,-1
\right),
\left(-3,1,-1\right),
\left(-3,-1,-1\right),
\nonumber \\
&\hspace{12pt}
\left(1,3,1\right),
\left(-1,3,1\right),
\left(1,-3,1\right),
\left(1,3,-1\right),
\nonumber \\
&\hspace{12pt}
\left(-1,-3,1\right),
\left(1,-3,-1\right),
\left(-1,3,-1\right),
\left(-1,-3,-1\right),
\nonumber \\
&\hspace{12pt}
\left(1,1,3\right),
\left(-1,1,3\right),
\left(1,-1,3\right),
\left(1,1,-3\right),
\nonumber \\
&\hspace{12pt}
\left(-1,-1,3\right),
\left(1,-1,-3\right),
\left(-1,1,-3\right),
\left(-1,-1,-3\right).
\end{align}
where we need to remove the ones 
not on the diamond lattice when implementing them into Eq.(1).  
\vspace{4mm}
\\
\noindent
{\it Helical and fan phases.} \\
The normalized vector spins of the classical helical and fan ground states are given by
%%%%%%%%%%%%%%%
\begin{align}
\bm{m}_{\bm{r}}
\propto
-\sin(\bm{q}\cdot\bm{r})
\bm{e}_{1\bar{1}0}
-
\cos(\bm{q}\cdot\bm{r}+\phi)
\bm{e}_{110}
+
M
\bm{e}_{\bm{h}}
,
\label{eq:mag1}
\end{align}
where $\phi=-\pi$ for the helical phase and $\phi=-3\pi/2$ for the fan phase,
$\bm{e}_{\mathrm{abc}}$ is the unit vector along the [abc] direction,
$\bm{e}_{\bm{h}}$ is the unit vector along the magnetic field $\bm{h}$,
and $\bm{q}=\frac{3\pi}{2}(1,1,0)$. 
The primitive translation vectors are
$\bm{a}_{1}=(2,2,0),
\bm{a}_{2}=\left(\frac{1}{2},-\frac{1}{2},0\right),
\bm{a}_{3}=(0,0,1)$, 
where there are 16 sites in the magnetic unit cell. 
The reciprocal lattice vectors are 
$\bm{b}_{1}=\frac{\pi}{2}(1,1,0),
\bm{b}_{2}=2\pi(1,-1,0),
\bm{b}_{3}=2\pi(0,0,1)$. 
\vspace{4mm}
\\
\noindent
{\it Antiferromagnetic skyrmion lattice phase}. \\
The normalized vector spins of the classical AFM-SkL ground state is given by 
%%%%%%%%%%%%%%%
\begin{align}
\bm{m}_{\bm{r}}
\propto
\sum_{m=1}^{3}
\left(
\sin(\bm{q}_{m}\cdot\bm{r})
\bm{e}_{m}
-\cos(\bm{q}_{m}\cdot\bm{r}-9\pi/8)
\bm{e}_{111}
\right)
+
M
\bm{e}_{\bm{h}}
,
\label{eq:mag2}
\end{align}
%%%%%%%%%%%%%%%
where $\bm{e}_{l}
=\bm{e}_{\bar{1}\bar{1}2},\;\bm{e}_{1\bar{2}1},\;\bm{e}_{2\bar{1}\bar{1}}$. 
and 
$\bm{q}_{1}\!=\!\frac{3\pi}{2}(1,-1,0),\:
\bm{q}_{2}\!=\!\frac{3\pi}{2}(1,0,-1),\:
\bm{q}_{3}\!=\!\frac{3\pi}{2}(0,1,-1)$ 
.
There are 384 sites in the magnetic unit cell spanned by 
$\bm{a}_{1}\!=\!(4,0,-4),\:
\bm{a}_{2}\!=\!(0,4,-4),\:
\bm{a}_{3}\!=\!(1,1,1)$.
The reciprocal lattice vectors are given by 
$\bm{b}_{1}\!=\!\frac{\pi}{6}(2,-1,-1),\:
\bm{b}_{2}\!=\!\frac{\pi}{6}(-1,2,-1),\:
\bm{b}_{3}\!=\!\frac{2\pi}{3}(1,1,1)$.
%%%%%
\vspace{4mm}
\\
\noindent
{\it Gauge transformation and spin wave Hamiltonian}. \\
For the spin wave analysis, we first introduce the three-dimensional rotation matrix as
%%%%%%%%%%%%%%%
\begin{align}
R_{\bm{r}}^{\mu\nu}=
n_{\bm{r}}^{\mu}n_{\bm{r}}^{\nu} +
(\delta^{\mu\nu}-n_{\bm{r}}^{\mu}n_{\bm{r}}^{\nu})\cos\phi_{\bm{r}}
-\sin\phi_{\bm{r}}\sum_{\rho}\epsilon^{\mu\nu\rho}
n_{\bm{r}}^{\rho},
\end{align}
%%%%%%%%%%%%%%%
where vector $\bm{n}_{\bm{r}}$ and angle $\phi_{\bm{r}}$ are defined as
%%%%%%%%%%%%%%%
\begin{align}
\bm{n}_{\bm{r}}
=
\frac
{\bm{m}_{\bm{r}}\times\bm{e}^{z}}
{|\bm{m}_{\bm{r}}\times\bm{e}^{z}|}
,
\hspace{20pt}
\phi_{\bm{r}}
=
\arccos(\bm{e}^{z}\cdot\bm{m}_{\bm{r}})
.
\end{align}
This matrix satisfies 
$R_{\bm{r}}\bm{m}_{\bm{r}}=\bm{e}^{z}$, 
meaning that it rotates the direction of the spin to the quantization axis ($z$-axis), 
which is the local gauge transformation 
changing the representation of the Hamiltonian but not the physical quantities. 
\par
We apply the Holstein-Primakoff transformation in the rotating frame, 
\begin{align}
R_{\bm{r}} \hat{\bm{S}}_{\bm{r}}
\simeq
\sqrt{\frac{S}{2}} (\hat{b}_{\bm{r}}+\hat{b}_{\bm{r}}^{\dagger})\bm{e}^{x}
-i \sqrt{\frac{S}{2}}(\hat{b}_{\bm{r}}-\hat{b}_{\bm{r}}^{\dagger})\bm{e}^{y}
+(S-\hat{b}_{\bm{r}}^{\dagger}\hat{b}_{\bm{r}})\bm{e}^{z}.
\end{align}

The spin Hamiltonian is approximated as $\hat{\mathcal{H}}\simeq E_{\mathrm{cl.}}+\hat{\mathcal{H}}_{\mathrm{mag.}}$,
where the magnon Hamiltonian $\hat{\mathcal{H}}_{\mathrm{mag.}}$ is calculated as
\begin{align}
\hat{\mathcal{H}}_{\mathrm{mag.}}
&=\sum_{\bm{r}}
\varepsilon_{\bm{r}} \hat{b}_{\bm{r}}^{\dagger} \hat{b}_{\bm{r}}
+ \frac{1}{2}\sum_{\bm{r}} \left(\lambda_{\bm{r}} (\hat{b}_{\bm{r}}^{\dagger})^{2} +
\mathrm{h.c.} \right)
\\
&+ \frac{1}{2}
\sum_{\bm{r}} \sum_{m} \sum_{\bm{\delta}_{m}} \left(
t_{\bm{r},\bm{\delta}_{m}} \hat{b}_{\bm{r}}^{\dagger} \hat{b}_{\bm{r}+\bm{\delta}_{m}}
+ \Delta_{\bm{r},\bm{\delta}_{m}}\hat{b}_{\bm{r}}^{\dagger}
 \hat{b}_{\bm{r}+\bm{\delta}_{m}}^{\dagger}X
+\mathrm{h.c.}\right).
\end{align}
%%%%%%%%%%%%%%%
We perform Fourier transformation 
$\hat{b}_{\bm{r}}=
\sqrt{N_s}^{-1} \sum_{\bm{k}}
\hat{b}_{\bm{k},\alpha} \mathrm{e}^{i\bm{k}\cdot\bm{r}}$,  
where $N_{s}=16$ and 384 for helical/fan and AFM-SkL phases, 
and $\alpha=1,2,\cdots,N_s$ is the sublattice site index within the magnetic unit cell.
%%%%%%%%%%%
\par
The magnon Hamiltonian is rewritten as using 
$\hat{\Phi}_{\bm{k}}=
(\hat{b}_{\bm{k},1},\cdots,\hat{b}_{\bm{k},N_{s}})$ as 
%%%%%%%%%%%%%%%
\begin{align}
\hat{\mathcal{H}}_{\mathrm{mag.}}
&=
\frac{1}{2}
\sum_{\bm{k}}
\hat{\Phi}_{\bm{k}}^{\dagger}
H_{\mathrm{BdG}}(\bm{k})
\hat{\Phi}_{\bm{k}}
+
\mathrm{const.}
, \\
&
H_{\mathrm{BdG}}(\bm{k})
=
\begin{pmatrix}
\Xi(\bm{k}) & \Delta(\bm{k})\\
\Delta^{*}(-\bm{k}) & \Xi^{*}(-\bm{k})
\end{pmatrix}
,
\end{align}
The magnon bands and eigenstates can be obtained by solving the eigenvalue equations, 
$\Sigma^{z} H_{\mathrm{BdG}}(\bm{k})
\bm{t}_{n}(\bm{k})=\varepsilon_{n}(\bm{k})
\bm{t}_{n}(\bm{k})$, 
with $\Sigma^{z}=\sigma^{z}\otimes I_{N_{s}\times N_{s}}$. 
By using magnon eigenstates, $\bm{t}_{n}(\bm{k})$, we obtain the Berry curvature 
and the thermal Hall conductivity in the main text.

%%%%%%%%%%%%%%%
%%%%%%%%%%%%%%%
%*%*%*%*%*%*%*%*%*%*%*%*%*%
\begin{figure*}[tbp]
   \centering
   \includegraphics[width=17cm]{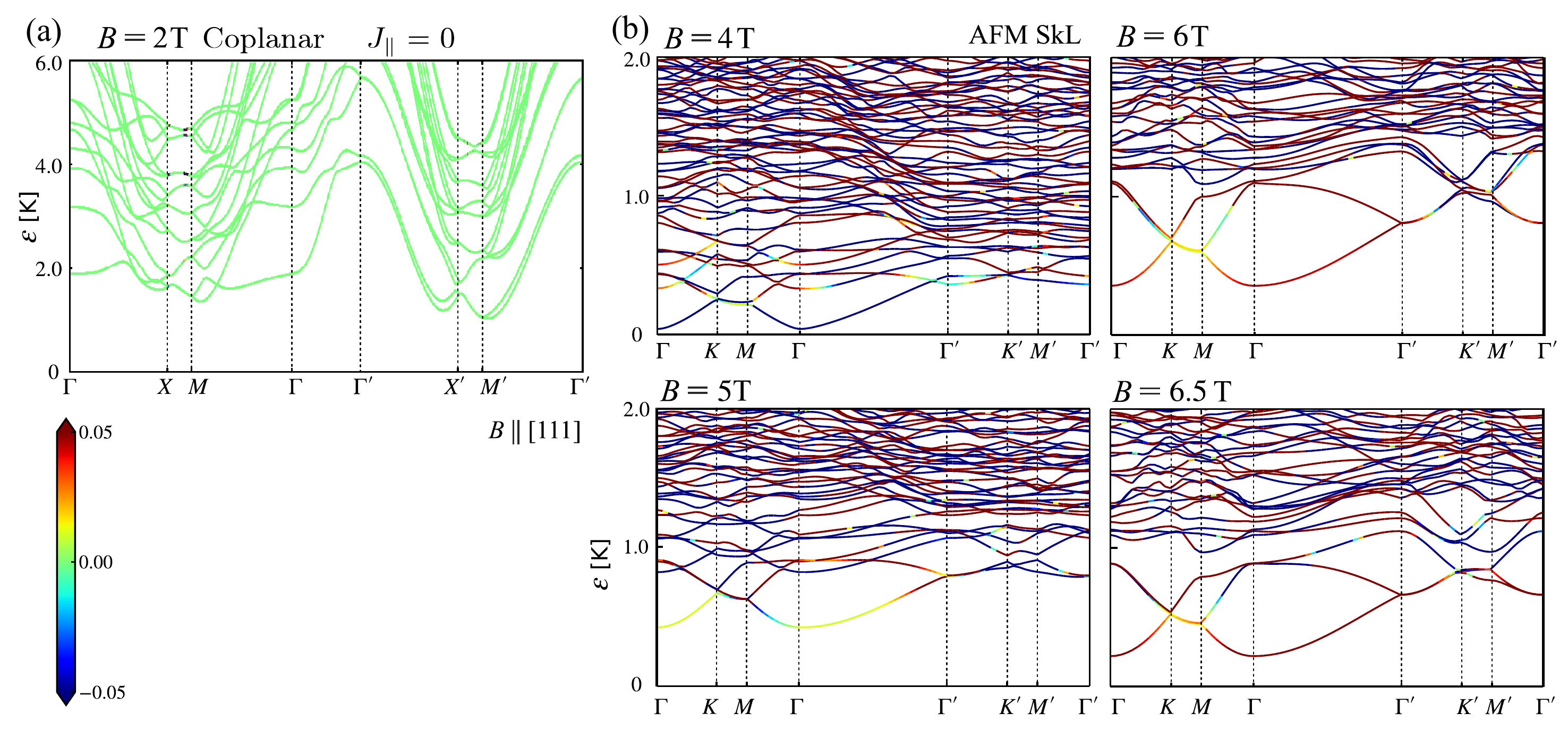}
   \caption{
Magnon bands obtained by the spin-wave theory in a field $\bm B\parallel [111]$. 
(a) Coplanar state at $B=2$\,T and $J_\parallel=0$ to be compared with 
the helical state in Fig.~\ref{fig2}(a), both with $N_c=16$. 
(b-d) AFM-SkL state with $N_c=384$ at $B=4, 5, 6, 6.5$\,T. 
Energy bands are shown in the color density plot of the Berry curvature $\Omega_{xy}^{(n)}$. 
The case of $\Omega_{xy}^{(n)}>0.05$ and $<-0.05$ 
are all plotted in red and blue, respectively. 
}
%\vspace{-4mm}
\label{fStheory}
\end{figure*}
%*%*%*%*%*%*%*%*%*%*%*%*%*%
%*%*%*%*%*%*%*%*%*%*%*%*%*%
\begin{figure*}[tbp]
   \centering
   \includegraphics[width=16.5cm]{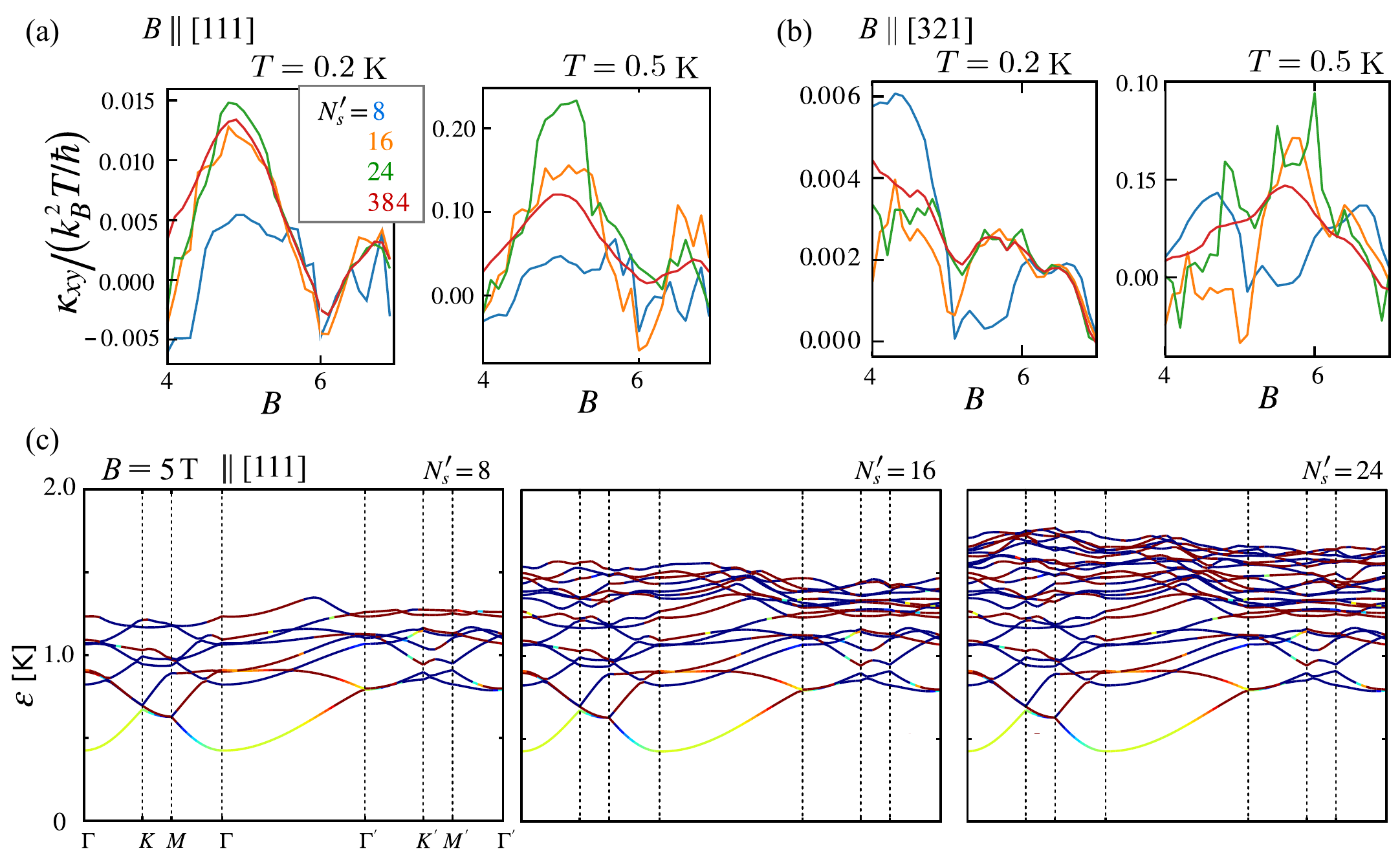}
   \caption{
(a,b) $\kappa_{xy}/(k_B^2T/\hbar)$ of the AFM-SkL phase 
obtained using the lowest $N_s'$-bands ($N_s'=8,16,24,384$) in taking the integral 
in 
for the field directions $\bm B\parallel [111]$ and $[321]$. 
(c) The description of lowest $N_s'$ bands ($N_s'=8,16,24$) at $B=5$\,T and $\bm B\parallel [111]$. 
}
\label{fStheory2}
\end{figure*}
%*%*%*%*%*%*%*%*%*%*%*%*%*%
\subsection{Variation of magnon bands}
\label{app:th-magband}
In Fig.~\ref{fStheory}(a) we show the magnon bands obtained at $B=2$\,T 
by artificially setting $J_\parallel=0$, which gives the coplanar magnetic structure 
instead of the helical one. 
There, we find that although the band structures look similar to the one 
shown in Fig.~\ref{fig2}(a) in the main text, they have mostly $\Omega^{n}_{xy}=0$, 
indicating that the coplanar phase does not have room to exhibit a thermal Hall effect. 
This indicates that the anisotropic exchange interactions are essential to have a finite Berry curvature. 
\par
Figure~\ref{fStheory}(b) shows the magnon bands for AFM-SkL phase in higher fields. 
At 5\,T, we find a much larger gap compared to the case of 4\,T, 
while for 6 and 6.5\,T the gap decreases. 
At the same time, we find tens of energy bands in the energy window of $\varepsilon \lesssim 2$\,K, 
while its number seems to decrease with increasing the field. 
This will influence the number of magnon carriers thermally excited, 
and explain the continuous decrease of $\kappa_{xx}$ in the experiment toward 8\,T. 
\par
For $\kappa_{xy}/T$, not only the number of low energy bands but 
the sign of $\Omega^{n}_{xy}$ matters, 
which cannot be visibly understood solely by the present plot, and will be examined 
in the next subsection. 

\subsection{Degree of contributions from magnon bands to $\kappa_{xy}$}
\label{app:th-kxy}
\noindent 
We now examine how the variation of energy bands influences $\kappa_{xy}$ 
by examining their contribution in detail. 
The Berry curvature $\Omega_{\mu\nu}^{(n)}(\bm{k})$ ($n=1,2,\cdots,N_s$) is 
described as
%%%%%%%%%%%%%%%
\begin{align}
\Omega_{\mu\nu}^{(n)}(\bm{k})
&=
-2
\mathrm{Im}
\left[
\frac{\partial\bm{t}_{n}^{\dagger}(\bm{k})}{\partial k_{\mu}}
\Sigma^{z}
\frac{\partial\bm{t}_{n}(\bm{k})}{\partial k_{\nu}}
\right]
\nonumber \\
&=
-2
\sum_{m=1}^{2N_{s}}
(1-\delta_{m,n})
(\Sigma^{z})_{m,m}
\nonumber \\
&\rule{0mm}{5mm}
\frac
{
\mathrm{Im}
\left[
\bm{t}_{n}^{\dagger}(\bm{k})
\frac{\partial H_{\mathrm{BdG}}(\bm{k})}{\partial k_{\mu}}
\bm{t}_{m}(\bm{k})
\bm{t}_{m}^{\dagger}(\bm{k})
\frac{\partial H_{\mathrm{BdG}}(\bm{k})}{\partial k_{\nu}}
\bm{t}_{n}(\bm{k})
\right]
}
{
(\varepsilon_{n}(\bm{k})
-
\varepsilon_{m}(\bm{k})
)^{2}+\delta
}
,
\label{eq:omega}
\end{align}
%%%%%%%%%%%%%%%
where $m=N_{s}+1,\cdots,2N_{s}$ denotes the particle-hole pairs. 
The first line in Eq.(\ref{eq:omega}) gives the expressions using only the information 
about the $n$-th band, while the second expression includes the inter-band matrix elements 
over the whole bands, making use of the completeness of the basis. 
We use the latter for numerical evaluation, 
where we introduce an infinitesimal positive number $\delta$ to reduce the numerical error 
due to divergence and take $\delta\rightarrow 0$. 
\par
We now test how the number of bands included in the calculation of $\kappa_{xy}$, 
denoted as $N_s'$ would influence the results. 
After obtaining the distribution of $\Omega_{xy}^{(n)}(\bm{k})$ over all the magnon bands, 
we confine the summation to the lowest $N_s'$-bands 
in the following formula, 
%%%%%%%%%%%%%%%
\begin{align}
\kappa_{xy}
=
-
\frac{k_{\mathrm{B}}^{2}T}{\hbar}
\int_{\mathrm{BZ}}
\frac{d^{3}\bm{k}}{(2\pi)^{3}}
\sum_{n=1}^{N_{s}'}
c_{2}[f(\varepsilon_{n}(\bm{k}))]
\Omega_{xy}^{(n)}(\bm{k})
.
\end{align}
%%%%%%%%%%%%%%%
Figure~\ref{fStheory2}(a) shows $\kappa_{xy}$ as function of $\bm B\parallel [111]$ for 
different choices of $N_s'=8,16,24$ and 384(full) at 
$T=0.2$ and 0.5\,K. 
At $T=0.2$\,K, the results converge already at $N_s'=16$, showing that 
only 10 bands or so contribute to $\kappa_{xy}$. 
However, at $T=0.5$\,K, the results differ much even when we increase as large as $N_s'=24$. 
The same tendency holds for $\bm B\parallel [321]$ shown in Fig.~\ref{fStheory2}(b). 
For clarification, we plot in Fig.~\ref{fStheory2}(c) the lowest $N_s'$-bands 
with  $N_s'=8,16,24$. 
One finds that already at $T=0.5$K, the magnon excitation following Bose statistics 
easily exceeds the energy window $\sim 2$\,K that accomodates $N_s'=24$. 
We thus clarify that in the AFM-SkL phase, the characteristic dense magnon energy bands 
contribute up to high energies in the thermal Hall effect. 
%*%*%*%*%**%*%*%*%*
%*%*%*%*%**%*%*%*%*
%*%*%*%*%**%*%*%*%*
%*%*%*%*%**%*%*%*%*
%*%*%*%*%*%*%*%*%*%*%*%*%*%
\begin{figure*}[tbp]
   \centering
  \includegraphics[width=16.5cm]{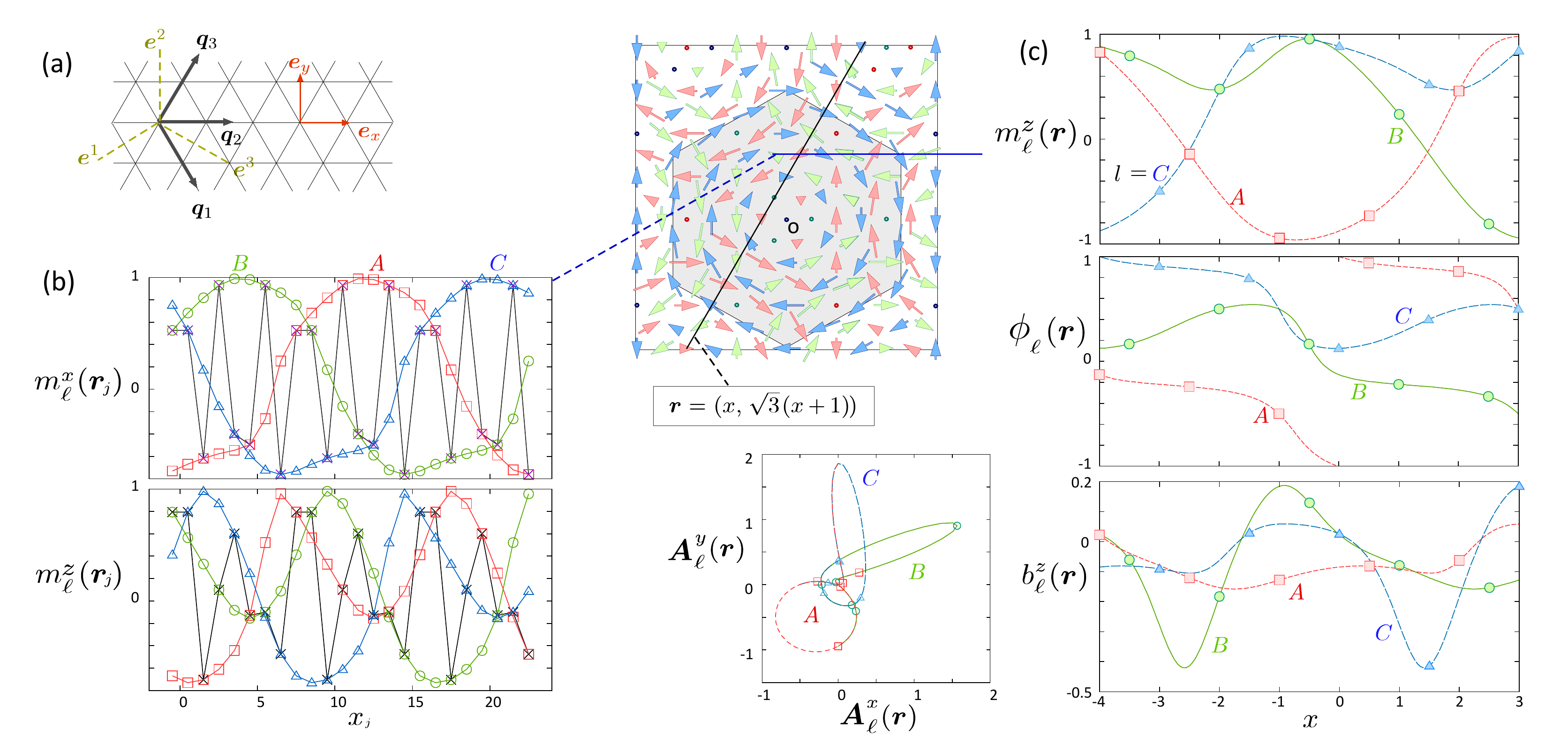}
   \caption{(a) Vectors defined in the two-dimensional $xy$-plane used in the formulation. 
(b) Magnetic structrues $m_\ell^x(\bm r_j)$ and $m_\ell^z(\bm r_j)$ obtained for discrete lattice sites 
$\bm r_j=(x_j,3\sqrt{3}/2)$. 
The one using Eq.(\ref{eq:mag}) give the magnetization for all three sublattices, 
whereas those in Eq.(\ref{eq:mag-abc}) give $\bm m_\ell$ of $\ell=A,B$, and $C$ sublattices separately, 
showing that the two equations are consistent. 
(c) The magnetization $m_\ell^z(\bm r)$, $\phi_\ell(\bm r)$, and $b_\ell^z(\bm r)$ 
along the $\bm r=(x,\sqrt{3}(x+1))$ line. 
The data points are the values for the discrete lattice points, and the lines are the 
obtained as continuous functions of $\bm r$. 
The corresponding $\theta_\ell(\bm r)$ and $m_\ell^x(\bm r) ,m_\ell^y(\bm r)$ are given in Fig.~\ref{fig4} in the main text. 
The variation of vector potential is given on the plane of $A_\ell^x(\bm r)$ and $A_\ell^y(\bm r)$. 
}
\label{fStheory3}
\end{figure*}
%*%*%*%*%*%*%*%*%*%*%*%*%*%
\subsection{Real-space spin texture and U(1) gauge fields in AFM-SkX on the triangular lattice}
\label{app:th-u1}
\noindent 
Here, we give the details of the calculations performed to obtain Fig.~\ref{fig4} in the main text. 
We consider the two-dimensional $xy$-plane that describes a single 
[111]-layer of MnSc$_{2}$S$_{4}$ that form the triangular lattice by taking [111] as $z$-axis. 
The orientations of the ordered magnetic moments of the AFM-SkL (see Fig.~\ref{fig1} and Fig.~\ref{fig4}(e) in the main text)
is given by the unit vector $\bm{m}(\bm{r})$ whose elements are given in the Cartesian coordinate as 
\begin{align}
\bm{m}(\bm{r})=&A^{-1} \Big(
\sum_{m=1}^{3}\big(-\sin(\bm{q}_{m}\cdot\bm{r})\bm{e}^{m}
+\cos\left(\bm{q}_{m}\cdot\bm{r}-\frac{9\pi}{8}\right)\bm{e}^{z}\big) \nonumber \\
& + M\bm{e}^{z}\Big)
\label{eq:mag}
\end{align}
where $A$ is a normalization factor to keep $|\bm m(\bm r)|=1$, 
$M$ is the spatially uniform magnetization due to a finite magnetic field, 
and $\bm{e}_{m}$ ($m=1,2,3$) are the unit vectors on the three-dimensional plane pointing in the 
direction perpendicular to $\bm{q}_{m}$ as
\vspace{-2mm}
\begin{align}
&\bm{e}^{1}=-\frac{\sqrt{3}}{2}\bm{e}^{x}-\frac{1}{2}\bm{e}^{y}, \hspace{5pt} 
\bm{e}^{2}=-\bm{e}^{y}, \hspace{5pt} 
\bm{e}^{3}=\frac{\sqrt{3}}{2}\bm{e}^{x}-\frac{1}{2}\bm{e}^{y}
\label{eq:e_ell}
\\
&\bm{q}_{1}=\frac{3\pi}{2}\big(\frac{1}{2}\bm{e}_{x}\!-\!\frac{\sqrt{3}}{2}\bm{e}_{y}\big), \hspace{5pt} 
 \bm{q}_{2}=\frac{3\pi}{2}\bm{e}_{x},\hspace{5pt}
 \bm{q}_{3}=\frac{3\pi}{2}\big(\frac{1}{2}\bm{e}_{x}\!+\!\frac{\sqrt{3}}{2}\bm{e}_{y}\big)
\end{align}
where we regard $\bm{e}^{\mu}$ and $\bm{e}_{\mu}$ as those defined 
in the spin space and in real space, respectively 
(see Fig.~\ref{fStheory3}(a)). 
\par
Let us consider the spatial variation of magnetic moments separately for the three sublattices 
$\ell =A,B,C$ to construct the U(1) gauge fields. 
This can be done by using the same $\bm{e}_{m}$ ($m=1,2,3$) as those given in Eq.(\ref{eq:e_ell}) 
but by taking different periods of propagation vectors 
\begin{align}
 \bm{Q}_{1}=\frac{\pi}{6}\big(\frac{1}{2}\bm{e}_{x}\!-\!\frac{\sqrt{3}}{2}\bm{e}_{y}\big), \hspace{5pt} 
 \bm{Q}_{2}=\frac{\pi}{6}\bm{e}_{x},\hspace{5pt} 
 \bm{Q}_{3}=\frac{\pi}{6}\big(\frac{1}{2}\bm{e}_{x}\!+\!\frac{\sqrt{3}}{2}\bm{e}_{y}\big)
\end{align}
and by shifting the origin depending on the sublattices to $\tilde{\bm r}_\ell=\bm r- \bm r_0^{\ell}$ 
with $\bm r_0^A=0$, $\bm r_0^B=(16,0)$, $\bm r_0^C=(8,0)$, 
we find 
\begin{align}
\bm{m}_\ell(\bm{r}) =& 
A^{-1}\Big( \sum_{m=1}^{3} 
   \big(-\sin(\bm{Q}_{m}\cdot \tilde{\bm r}_\ell)\bm{e}^{m}
      +\cos\left(\bm{Q}_{m}\cdot \tilde{\bm r}_\ell   -\frac{9\pi}{8}\right)\bm{e}^{z}\big)\nonumber\\
   & + M\bm{e}^{z}\Big)
\label{eq:mag-abc}
\end{align}
Since $\bm{m}_\ell(\bm{r})$ is described using $(\theta_\ell(\bm r), \phi_\ell(\bm r))$, 
we can rewrite these elements as 
\begin{align}
\sin\theta_\ell(\bm{r})\cos\phi_\ell(\bm{r}) 
&=\frac{\sqrt{3}}{2A}
\Big(\sin(\bm{Q}_{1}\cdot \tilde{\bm r}_\ell)-\sin(\bm{Q}_{3}\cdot\tilde{\bm r}_\ell)\Big)
\nonumber\\
\sin\theta_\ell(\bm{r})\sin\phi_\ell(\bm{r}) 
  &=\frac{-1}{2A}\Big(\sin(\bm{Q}_{1}\cdot\tilde{\bm r}_\ell)
  +2\sin(\bm{Q}_{2}\cdot\tilde{\bm r}_\ell)+\sin(\bm{Q}_{3}\cdot\tilde{\bm r}_\ell)\Big)
\nonumber\\
\cos\theta_\ell(\bm{r}) &= \frac{1}{A}\Big(\sum_{m}\cos\left(\bm{Q}_{m}\cdot\tilde{\bm r}_\ell
 -\frac{9\pi}{8}\right)+M\Big), 
\end{align}
which gives the description of magnetic moments in the continuous space $\bm r$ in the $ xy$ plane. 
Figure~\ref{fStheory3}(b) shows the comparison of 
$\bm m(\bm r)$ in Eq.(\ref{eq:mag}) and 
$\bm{m}_\ell(\bm{r})$ in Eq.(\ref{eq:mag-abc}) along the $\bm r_j=(x_j,0)$ line 
calculated for discrete lattice points 
where we set $M=0$. 
At each lattice point, they cross while the spatial periods for the continuous $\bm r$ differ. 
This shows that the two equations are consistent with each other. 
\par
The U(1) gauge field is constructed using slowly varying sets of $\bm{m}_\ell(\bm{r})$. 
The vector potential for the three sublattices is given as 
\begin{equation}
\bm A_\ell(\bm r)= -\frac{\cos\phi_\ell(\bm r)}{\tan\theta_\ell(\bm r)}\nabla \bm m_\ell^y(\bm r)
+ \frac{\sin\phi_\ell(\bm r)}{\tan\theta_\ell(\bm r)}\nabla \bm m_\ell^x(\bm r), 
\end{equation}
and the fictitious magnetic field generated as a rotation of the vector field $\bm A_\ell(\bm r)$ 
is given as 
\begin{equation}
b^z_\ell(\bm r)= \partial_x A_\ell^y(\bm r)-\partial_y A_\ell^x(\bm r)
= \bm m_\ell(\bm r)\cdot \big(\partial_x \bm m_\ell(\bm r) \times \partial_y \bm m_\ell(\bm r)\big), 
\end{equation}
where  
\begin{align}
\partial_\mu \bm m_\ell^x(\bm r)
=&\frac{\sqrt{3}}{2A}
\Big((\bm Q_1)_\mu \cos(\bm{Q}_{1}\cdot \tilde{\bm r}_\ell)-
(\bm Q_3)_\mu\cos(\bm{Q}_{3}\cdot\tilde{\bm r}_\ell)\Big),
\nonumber\\
\partial_\mu \bm m_\ell^y(\bm r)
  =&\frac{1}{2A}\Big(
    (\bm Q_1)_\mu \cos(\bm{Q}_{1}\cdot\tilde{\bm r}_\ell)
  +2 (\bm Q_2)_\mu \cos(\bm{Q}_{2}\cdot\tilde{\bm r}_\ell) \nonumber \\
 &  + (\bm Q_3)_\mu \cos(\bm{Q}_{3}\cdot\tilde{\bm r}_\ell) \Big),
\nonumber\\
\partial_\mu \bm m_\ell^z(\bm r) =& -\frac{1}{A}\big[
  \sum_{m} (\bm Q_m)_\mu \sin\big(\bm{Q}_{m}\cdot\tilde{\bm r}_\ell
 -\frac{9\pi}{8}\big)\big]. 
\end{align}
Figure~\ref{fStheory3} (c) shows $m_\ell^z(\bm r)$, $\phi_\ell(\bm r)$ and $b_\ell^z(\bm r)$ 
separately for the three sublattices. 
They are the continuous functions of $\bm r$ where we put the symbols at the discrete lattice points. 
The corresponding data of $\theta_\ell(\bm r)$ and $m_\ell^x(\bm r)$-$m_\ell^y(\bm r)$ profile are shown 
in Fig.~\ref{fig4} in the main text.

%%%%%%%%%%%%%%%%%%%%%%%%%%%%%%%%%%%%%%%%%%%%
%%%%%%%%%%%%%%% bibliography %%%%%%%%%%%%%%%
%%%%%%%%%%%%%%%%%%%%%%%%%%%%%%%%%%%%%%%%%%%%
\bibliographystyle{naturemag}
\bibliography{mnsc2s4ref}
%%%%%%%%%%%%%%%%%%%%%%%%%%%%%%%%%%%%%
\end{document}